\documentclass[12pt]{iopart}

\usepackage{graphicx}
\begin{document}

\title[Strong coupling theory of heavy fermion criticality II]{Strong coupling theory of heavy fermion criticality II}

\author{Peter W\"{o}lfle}

\address{Institute for Theory of
Condensed Matter, Karlsruhe Institute of Technology, 76049 Karlsruhe,
Germany}
\address{Institute for Nanotechnology, Karlsruhe Institute of Technology, 76031 Karlsruhe, Germany}
\ead{peter.woelfle@kit.edu}

\author{J\"{o}rg Schmalian}

\address{Institute for Theory of
Condensed Matter, Karlsruhe Institute of Technology, 76049 Karlsruhe,
Germany}
\address{Institute for Solid State Physics, Karlsruhe Institute of Technology, 76031 Karlsruhe,
Germany}
\ead{joerg.schmalian@kit.edu}
\author{Elihu Abrahams}

\address{Department of Physics and Astronomy, University of California
Los Angeles, Los Angeles, CA 90095}
\ead{abrahams@physics.ucla.edu}

\begin{indented}
\item[]\today
\end{indented}

\begin{abstract}
We present a theory of the scaling behavior of the
thermodynamic, transport and dynamical properties of a three-dimensional
metal governed by $d$-dimensional fluctuations at a quantum critical point, where the
electron quasiparticle effective mass diverges. We determine how the
critical bosonic order parameter fluctuations are affected by the
effective mass divergence. The coupled system of fermions and bosons is
found to be governed by two stable fixed points: the conventional
weak-coupling fixed point and a new strong-coupling fixed point, provided
the boson-boson interaction is irrelevant. The latter fixed point supports
hyperscaling, characterized by fractional exponents. The theory is applied to
the antiferromagnetic critical point in certain heavy fermion compounds, in which
the strong-coupling regime is reached.
\end{abstract}

%
%
%
%
%

\section{Introduction}

The properties of matter at low temperatures has always been a
subject of frontier research in theoretical and experimental condensed
matter physics. In the case of metallic systems, in particular, this
research has led to an appreciation of the crucial role played by
electron-electron interaction effects in determining the character of
unexpected new phases that emerge at low temperature. Thus, in recent
decades, there have been intensive studies of quantum criticality in metals
and magnetic materials. A zero-temperature transition between two distinct
ground states, a quantum phase transition, may be realized by tuning some
external parameter such as magnetic field, pressure, chemical
composition,\ldots Recent metallic materials realizations of quantum phase
transitions include the cuprate superconductors, the iron-based
superconductors and the heavy-electron compounds. All these have complex
phase diagrams that reflect how strong electron correlation effects
determine the character and competition of new phases (\textit{e.g.}
magnetism, superconductivity) and behavior that deviates from that of
simpler metals, as described by the standard Landau Fermi liquid theory.

Often, a quantum critical point (QCP) marks the transition from
a magnetically-ordered phase to a paramagnetic one; associated with it are
bosonic quantum fluctuations of the relevant magnetic order parameter. Then,
due to the scattering of the fermionic quasiparticles off these
fluctuations, there may be significant modifications of Fermi liquid
behavior even in the paramagnetic regions of the phase diagram. Quantum
phase transitions in itinerant magnetic systems have been theoretically
studied using approaches related to that originally proposed by Hertz\cite%
{JH} and further developed by Millis\cite{millis} and these have been recently reviewed.\cite{hvl} The early theories of
quantum critical behavior were formulated in the framework of a $\phi ^{4}$%
-field theory defined by a Ginzburg-Landau-Wilson action of an order
parameter field $\phi$; they found that since temporal fluctuations increase
the effective dimension of the field theory to $d_{eff}=d+z$ where $d,z$ are
the spatial dimension of the fluctuations and the dynamical critical
exponent, it may happen that the theory is essentially Gaussian. This is the
case if $d_{eff}>4$, the upper critical dimension, and the fluctuations are
effectively non-interacting (for a review see Ref.~\cite{hvl}). The
theory for non-metallic systems expressed in terms of a field theory for the
order parameter is well-founded; it was developed in Ref.~\cite{CHN}.
However, in metallic systems, the fermionic degrees of freedom may become
critical themselves and a generalized formulation that treats the critical
bosons and fermions on equal footing is called for. The large
low-temperature quasiparticle mass enhancements observed in heavy-electron
compounds suggest that they are systems in which this situation may occur.
It is in this context that a semi-phenomenological theory for quantum
criticality was developed \cite{qp1} and applied\cite{pnas,asw} to successfully  account for
experimental observations in the two compounds YbRh$_2$Si$_2$ (YRS)
and CeCu$_{6-x}$Au$_x$ (CCA). In this report, we review the fundamentals of that
theory of ``critical quasiparticles", fill in some of the mathematical
details and put it on firmer microscopic grounds by deriving the scaling
behavior of the thermodynamic, transport and dynamical properties of the
metallic quantum critical system.

Contrary to what one might expect, there actually exists a
window in parameter space where a consistent theory may be constructed,
although the fermionic and bosonic degrees of freedom are strongly coupled.
A necessary condition to be fulfilled is, however, that the direct
boson-boson coupling is still irrelevant (crudely speaking, $d_{eff}>4$).
Our theory presents an alternative scenario of quantum criticality in heavy-fermion metals. 
In these compounds, the accepted scenario is that the Kondo effect leads to the emergence of
heavy quasiparticle masses as a consequence of the screening of the localized spins of the rare earth ions  by the conduction electrons. This leads to a many-body resonance
state at each ion, at which the conduction electrons are resonantly
scattered, which slows down their motion through the lattice. On the other
hand, the localized spins are coupled by exchange interaction, mediated by the
conduction-electron system (RKKY) or by direct exchange. As pointed out
early on by Doniach,\cite{Doniach77} the competition between exchange and the  Kondo effect
might lead to an abrupt breakdown of Kondo screening, often called ``Kondo
breakdown". Following the experimental discovery of unusual quantum critical
behavior in many heavy-fermion compounds, several ``Kondo breakdown"
scenarios have been proposed.\cite{Si10} One of these 
involves a local transition of the Kondo resonance state at the
antiferromagnetic transition, giving rise to ``local quantum criticality". 
A somewhat similar proposal invokes an ``orbitally
selective Mott transition" involving the abrupt vanishing of the
hybridization of $f$ electrons and conduction electrons within an Anderson
lattice model, and hence a sharp transition between the itinerant heavy-fermion state and a localized $f$-electron spin state. This picture has been
realized using slave-boson mean field theory \cite{Pepin07} and
cellular-DMFT calculations.\cite{Kotliar08} Yet another scenario proposes a
fractionalization of the heavy quasiparticles into spin and charge carrying
components driven by strong frustration and/or quantum fluctuations,\cite{Senthil06} 
leading to the formation of a
spin liquid state \cite{Sachdev08,Lee08} dubbed ``fractionalized Fermi
liquid"  or FL*. The latter may undergo a transition into a usual antiferromagnetic
phase. A third possibility is given by a Fermi surface reconstruction inside
the antiferromagnetically ordered phase, as the hybridization of $f$-electrons
and conduction electrons is tuned through a critical value, changing the
large Fermi surface of the heavy quasiparticles into the small Fermi surface
of light conduction electrons. The latter has apparently been observed in
some of the CeMIn$_5$ compounds (M stands for a transition metal ion).\cite{Park08,Shishido05,Goh08} It has also been recognized that the two phases, heavy quasiparticles in the paramagnetic
phase and localized ordered spins weakly coupled to conduction electrons may
be adiabatically connected.\cite{Vojta08} An adiabatic connection of states
characterized by large and small Fermi surfaces is indeed possible in the
antiferromagnetic phase, where the Fermi surfaces are downfolded into the
magnetic Brillouin zone and the meaning of large and small Fermi surface may
be lost. We assume here that the latter scenario is applicable, {\it i.e.} one has
a smooth crossover from large heavy-fermion Fermi surface to ordered local
moments, as one moves deeper into the antiferromagnetic phase. This
assumption appears justified in the case of the two heavy fermion compounds on which we focus, CeCu$_{6-x}$Au$_{x}$ at $x\approx 0.1$ and YbRh$_{2}$Si$_{2}$. For the latter, YRS, various anomalies were observed  along  a $T^*$-line in the temperature-magnetic field phase diagram.\cite{gegen07} These were initially interpreted as a consequence of ``Kondo breakdown." However, ARPES studies \cite{Kummer15} do not give an indication for a Fermi
surface reconstruction. Recently, the anomalies observed along the $T^*$-line have been shown to arise from a freezing out of spin-flip scattering.\cite{WA15}

\section{Critical Fermi liquid}

The unusual properties of metals  observed near quantum
critical points have often been termed ``non-Fermi liquid behavior". 
Rather than describing such a state as what it is not, we shall call it a
``critical Fermi liquid" and its low-lying excitations ``critical
quasiparticles". While it is true that at zero temperature $T=0$ and zero
excitation energy $\omega =0$ fermionic quasiparticles may no longer exist,
we shall demonstrate that at any non-zero $T,\omega$, fermionic
quasiparticles are still well-defined for a class of critical states. As
first introduced by Landau, fermionic quasiparticles are defined by the
poles of the single particle Green's function $G(\mathbf{k},\omega )$ in the
complex frequency plane. The retarded Green's function is given in terms of
the self energy $\Sigma (\mathbf{k},\omega )$ as 
\begin{equation}
G_{\sigma }(\mathbf{k},\omega +i0)=\frac{1}{\omega -\epsilon _{\mathbf{k}%
}-\Sigma _{\sigma }(\mathbf{k},\omega +i0)}
\end{equation}%
where $\sigma $ is the spin quantum number. For the moment, we assume that
the dependence of $\Sigma$ on ${\mathbf{k}}$ is weak and may be neglected.
We shall return to this issue later. When the Green's function has a
(quasiparticle) pole at $\omega = \epsilon_{\mathbf{k}}^*$, we may expand $%
\Sigma$ near it as  $\Sigma _{\sigma }(\mathbf{k},\omega +i0)=\mathrm{Re}%
\Sigma _{\sigma }(\mathbf{k},\epsilon _{\mathbf{k\sigma }}^{\ast })+i\mathrm{%
Im}\Sigma _{\sigma }(\mathbf{k},\epsilon _{\mathbf{k\sigma }}^{\ast
})+(\omega -\epsilon _{\mathbf{k\sigma }}^{\ast })\partial \mathrm{{Re}%
\Sigma _{\sigma }(\mathbf{k},\omega )/\partial \omega |_{\omega =\epsilon _{%
\mathbf{k\sigma }}^{\ast }}+...}$ such that the Green's function may be
expressed in terms of a quasiparticle term and and an incoherent background
contribution: $G_\sigma= G_\sigma^{qp}+G_\sigma^{inc}$, where 
\begin{equation}
G_{\sigma }^{qp}(\mathbf{k},\omega +i0) =\frac{Z_{\sigma }(\mathbf{k}%
,\epsilon _{\mathbf{k\sigma }}^{\ast })}{\omega -\epsilon _{\mathbf{k\sigma }%
}^{\ast }+i\Gamma _{\sigma }(\mathbf{k},\epsilon _{\mathbf{k\sigma }}^{\ast
})}
\end{equation}%
with $[Z_{\sigma }(\mathbf{k},\epsilon _{\mathbf{k\sigma }}^{\ast
})]^{-1}=1-\partial \mathrm{{Re}\Sigma _{\sigma }(\mathbf{k},\omega
)/\partial \omega |_{\omega =\epsilon _{\mathbf{k\sigma }}^{\ast }}}$ and $%
\Gamma _{\sigma }(\mathbf{k},\epsilon _{\mathbf{k\sigma }}^{\ast
})=Z_{\sigma }(\mathbf{k},\epsilon _{\mathbf{k\sigma }}^{\ast })\mathrm{Im}%
\Sigma _{\sigma }(\mathbf{k},\epsilon _{\mathbf{k\sigma }}^{\ast })$. 
The quasiparticle contribution to the spectral function $\mathcal{A}%
^{qp}=(-1/\pi) \mathrm{Im}\,G^{qp}$ is then 
\begin{equation}
\mathcal{A}^{qp}_\sigma(\mathbf{k},\omega)=\frac{Z _\sigma(\mathbf{k}%
,\epsilon_{\mathbf{k}}^*) \Gamma_\sigma(\mathbf{k},\epsilon_{\mathbf{k}%
}^*)/\pi}{[\omega - \epsilon_{\mathbf{k}}^*]^2 +\Gamma_\sigma(\mathbf{k}%
,\epsilon_{\mathbf{k}}^*)^2}  \label{qpeq}
\end{equation}
It is seen that $\mathcal{A}(\omega)$ has a quasiparticle peak at $\omega=
\epsilon_{\mathbf{k}}^*$, provided $\Gamma <\epsilon_{\mathbf{k}}^*$. In
that case, $\mathcal{A}(\omega)\approx Z(\omega)\delta(\omega - \epsilon_{%
\mathbf{k}}^*)$, we may replace $\epsilon_{\mathbf{k}}^*$ by $\omega$ and
observe that $\epsilon_{\mathbf{k}}^*\approx Z\epsilon_{\mathbf{k}} $ so
that we may interpret $1 /Z$ as a correlation-induced mass enhancement $m^*/m
$.
We now explore the possibility of using the notion of fermionic
quasiparticles even when the quasiparticle weight $Z_{\sigma }(\mathbf{k}%
,\omega )\rightarrow 0$ as $\omega \rightarrow 0$. As long as $\omega \neq 0$
(or else $\omega =0$, but $T\neq 0$), so that $Z\neq 0$, what we defined as
the quasiparticle contribution to $G$ does exist. The question is whether it
is well-defined. We have seen that a quasiparticle peak in the spectral
function exists if the decay rate $\Gamma (\omega)$ (the width of the peak)
is less than the energy at the peak position, 
Consider the situation of a self energy that varies with frequency  as
\begin{equation}
\Sigma(\omega)\propto i(i\omega)^{1-\eta}= i(\cos\phi +i\sin\phi)|\omega|^{1-\eta},\;\;\;\; 0<1-\eta<1, 
\end{equation}
where $\phi ={\rm sign}(\omega)(1-\eta)\pi/2$. We shall see that this behavior is actually found
near a QCP. This form is chosen such that $\mathrm{Im}\Sigma >0$, but $%
\mathrm{{Re}\Sigma \propto sign(\omega )}$, as required by Fermi statistics
and the analyticity properties of $\Sigma$. We note that $\mathrm{{Re}%
\Sigma \propto Im\Sigma \propto |\omega |^{1-\eta }}$.\ Here we have
suppressed the momentum dependence of $\Sigma$ for simplicity. It follows that at low frequency,
$Z(\omega )\propto |\sin \phi |^{-1}(1-\eta )^{-1}|\omega |^{\eta }$, so that the quasiparticle decay rate $\Gamma (\omega )= Z(\omega){\rm Im}\Sigma(\omega)\propto (1-\eta
)^{-1}\cot [(1-\eta )\pi /2]|\omega |$ is a linear function of $\omega$
independent of $\eta.$ From these considerations, we may deduce the condition on the exponent $\eta$ such that coherent quasiparticles  exist, {\it i.e.} $\Gamma(\omega)<\omega$. We see that it is required that
\begin{equation}
\cot [(1-\eta )\pi/2]<(1-\eta)   \rightarrow  \eta < 0.36.
\label{eta}
\end{equation}
We shall see that this condition has important consequences when the self energy varies around the Fermi surface.

When $Z\propto $ $|\omega |^{\eta }$, the quasiparticle
effective mass will be scale-dependent, $m^{\ast }=m^{\ast }(\omega )\propto
Z^{-1}\propto |\omega |^{-\eta }$. The effective mass is a measure of the
scale dependent quasiparticle density of states $N^{\ast }(\omega )\propto
m^{\ast }(\omega ) $, which enters the specific heat coefficient $\gamma
(T)=C(T)/T\propto N^{\ast }(T)$. So whenever the experimentally determined $%
\gamma (T)$ appears to diverge in the neighborhood of a QCP, we may assume
the fermionic quasiparticles to be critical.

Whether or not the momentum dependence of the self energy may
be neglected, as assumed above, depends on the microscopic processes that
contribute to it. As mentioned in Sec.~I, this is often the scattering of
the quasiparticles off quantum critical fluctuations of the order parameter.
In the case of incommensurate antiferromagnetic or charge density wave
order, these fluctuations are at some wavevector(s) $\mathbf{Q}$ that
connect discrete sets of points on the Fermi surface, the so-called ``hot
spots". For quasiparticles near the hot spots, the momentum dependence of $%
\Sigma$ may not be neglected. We will take this dependence into account in
the following.

\section{Renormalized perturbation theory of self energy
and vertex function.}

The critical spin fluctuations near a quantum critical point
separating paramagnetic and antiferromagnetic phases are described by the
dynamical spin susceptibility $\chi (\mathbf{q},\nu )$ at wave vectors $%
\mathbf{q}$ near the ordering wave vector $\mathbf{Q}$. We assume
incommensurate antiferromagnetic order, i.e. $\mathbf{Q}$ is not a
reciprocal lattice vector, as is most often the case in metallic
antiferromagnets. The generic form of $\chi (\mathbf{q},\nu )$ is 

\begin{equation}
\chi({\bf q},\nu) = \frac{N_0}{r +({\bf q-Q})^2\xi_0^2 - i\Lambda_Q^2(\nu/v_FQ)}
\label{X}
\end{equation}
where $N_{0}$ is the bare density of
states at the Fermi level, $r$ is the control parameter tuning the system
through the QCP, $\xi _{0}\approx k_{F}^{-1}$ is a microscopic correlation
length, $v_{F}=k_{F}/m$ is the bare Fermi velocity, and $\Lambda _{Q}=\Lambda (\mathbf{k},\omega =0;\mathbf{q}%
,\nu )$ is the vertex function at the antiferromagnetic
spin fluctuation-particle-hole vertex, {\it i.e.} the vertex at non-zero momentum $%
{\bf q}\approx {\bf Q}$. The physical (renormalized) susceptibility is determined by the structure and dynamics of the underlying electron degrees of freedom. This is reflected in Eq.~({\ref X}) by the appearance of
$\Lambda _{Q}$ in the Landau damping term in the denominator. It may be
shown that when $Z^{-1}(\omega )\propto m^{\ast }(\omega )/m$ diverges, then
the vertex $\Lambda _{Q} \sim Z^{-1}$ will diverge as well
(Ref.~\cite{WA16} and below). The dynamical
properties (self energy, vertex function) of the critical Fermi liquid will
be determined by the scattering of the quasiparticles from the renormalized
critical fluctuations. It follows that $\chi({\bf q},\nu)$ depends on
the critically enhanced effective mass. Likewise, the effective mass
enhancement following from scattering of quasiparticles off spin
fluctuations depends on the spectral density of spin fluctuations. The
resulting set of self-consistent equations for the effective mass and the
vertex function supports new strong coupling solutions as will be shown
below. 

As will be reviewed in the next subsection, the self energy and
the vertex functions are strongly dependent on the position on the Fermi
surface. In particular, for scattering by a single spin fluctuation,
involving the large momentum transfer $\mathbf{Q}$, the hot spots consist of
a limited manifold of $\mathbf{k}$-states on the Fermi surface. This
manifold comprises those wavevectors $\mathbf{k}_h$ that satisfy the
condition that $\mathbf{k}_{h}$ and $\mathbf{k}_{h}\pm\mathbf{Q}$ are both
near the Fermi surface. 
Landau damping involves the decay of spin fluctuations of frequency $\nu$
and momentum $\mathbf{q}$ into particle-hole pairs of momenta $\mathbf{k+q}$
and $\mathbf{k}$ close to the Fermi surface, within an energy range of order 
$\max \{\nu ,T\}$. At momenta $\mathbf{q}$ sufficiently close to the
ordering momentum $\mathbf{Q}$ these particle-hole pairs will be in the hot
spots. Below we will estimate the extension of the hot spots on the Fermi
surface as $|\delta\mathbf{k}_h| \approx k_F|\nu/\epsilon_F|^{1/2}\Lambda_Q$%
. When $\mathbf{q}$ deviates from $\mathbf{Q}$ by more than $|\delta\mathbf{k%
}_h|$, the particle hole pairs involved in the Landau damping will be
outside the hot spots. In the calculation of the self energy, which follows in the next section, it  will be seen that 
the wavevectors of the
relevant fluctuations contributing to the self energy satisfy exactly this
condition, namely $|{\bf q}-{\bf Q}|\gg\delta k_h \sim \nu^{1/2}\Lambda_Q$. We therefore conclude that the vertex function $\Lambda _{Q}$
appearing in the Landau damping is actually the one for the cold
quasiparticles.

\subsection{Self Energy $\Sigma $}

\subsubsection{One-loop approximation}

In one-loop approximation, \textit{i.e.} scattering from a
single spin fluctuation, the imaginary part of the self-energy for a ``hot"
quasiparticle is given by

\begin{eqnarray}
\mathrm{Im}\Sigma ^{(a)}(\mathbf{k},\omega ) &=&u^{2}\Lambda _{Q,h}^{2}\int 
\frac{d\nu }{2\pi }\int \frac{d^{d}q}{(2\pi )^{d}}\mathrm{Im}G(\mathbf{k+q}%
,\omega +\nu)  \nonumber \\
&&\times \mathrm{Im}\chi (\mathbf{q},\nu )[b(\nu )+f(\omega +\nu )], \label{1}
\end{eqnarray}%
where $b(\nu )=1/(e^{\nu /T}-1)$ and $f(\omega )=1/(e^{\omega /T}+1)$ are
the Bose and Fermi functions, $u\approx N_{0}^{-1}$ is the boson-fermion
interaction and $\Lambda _{Q,h}$ is the relevant vertex function (see
above). In the limit of low excitation energy, $\omega \rightarrow 0$\ and
at the Fermi surface, $\epsilon _{\mathbf{k}}=0$, we take into account that
the main contribution to the integrals is from small $\nu $, so we may
approximate $\mathrm{Im}G$ by its quasiparticle component, $\mathrm{Im}%
G^{qp}(\mathbf{q+k},\nu +\omega )\approx \pi Z\delta (\omega +\nu -\epsilon
_{\mathbf{k+q}}^{\ast })$, as already mentioned below Eq.~(\ref{qpeq}). We
now shift the momentum vector $\mathbf{q\rightarrow q+Q}$ , such that the
peak of $\chi $ shifts to $\mathbf{q=}0$ and $\epsilon_{\mathbf{k +q}}^* \to
\epsilon_{\mathbf{k +q+Q}}^*$. We expand the quasiparticle energy $\epsilon^*
$ in both small $q$ and small deviation $\delta\mathbf{k}$ from the hot spot
momentum, so that $\epsilon^*_{\mathbf{k}_h+\mathbf{Q+q}+\delta\mathbf{k}%
}\approx  q|v^*_{\mathbf{k}_h +\mathbf{Q}}|\cos\theta + \delta\epsilon^*_{%
\mathbf{k}}$, where $\delta\epsilon^*_{\mathbf{k}}= \delta\mathbf{k}\cdot%
\mathbf{v}^*_{\mathbf{k}_h +\mathbf{Q}}$ + O$(\delta\mathbf{k}^2)$ and $%
\theta $ is the angle subtended by the vectors $\mathbf{q}$ and $\delta%
\mathbf{k}$. We have set $\epsilon _{\mathbf{k}_{h}\mathbf{+Q}}^{\ast }=0$,
since $\mathbf{k}_h+\mathbf{Q}$ is on the Fermi surface. We perform the
angular integral:%
\begin{equation}
\int d\Omega _{q}\mathrm{Im}G^{qp}(\mathbf{k}_{h}\mathbf{+q},\omega +\nu )\propto 
\frac{1}{v_{F}q}\Theta (qv_{F}^{\ast }-|\delta \epsilon _{\mathbf{k}}^{\ast
}|),  \label{theta}
\end{equation}%
where $\Theta (x)$ is the unit step function.\ In the limit of low $%
T\ll\omega $ the $\nu $-integral extends over the interval $[0,\omega ]$. The 
$q$-integral is governed by the lower limit $q_{l}\approx \lbrack (|\nu
|\Lambda _{Q}^{2})^{2}+(\delta \mathbf{k})^{4}]^{1/4}$, as determined by the
theta function of Eq.~(\ref{theta}) and the denominator of $\mathrm{Im}\chi(%
\mathbf{q},\nu)$ from Eq.~({\ref{X}), where we set $v_{F}Q\approx \epsilon
_{F}$ (as $Q$ is usually of order $k_F$). Here and in what follows we have
set $\xi_0\approx 1/k_F$ and used $k_F,\epsilon_F$ as units for wave vectors
and energies or frequencies. Then 
\begin{eqnarray}
\mathrm{Im}\Sigma ^{(a)}(\mathbf{k},\omega ) &\propto& u^{2}\Lambda
_{Q,h}^{2}N_{0}\xi _{0}^{-4}v_{F}^{-1}\int_{0}^{\omega }d\nu \frac{\nu }{%
\epsilon _{F}}\Lambda _{Q}^{2}\int_{q_{l}}\frac{dq}{q^{6-d}}  \nonumber \\
&\propto& \Lambda _{Q,h}^{2}\Lambda _{Q}^{d-3}\epsilon _{F}\{[\omega ^{2}+x_{%
\mathbf{k}}^{2}]^{(d-1)/4}-x_{\mathbf{k}}^{(d-1)/2}\}, \label{Sigma_k}
\end{eqnarray}%
where $x_{\mathbf{k}}=(\delta\mathbf{k} /\Lambda _Q)^2$.  Therefore
\begin{equation}
{\rm Im}\Sigma^{(a)}({\bf k}, \omega) 
\propto
\left\{ \begin{array}{cc}
 \Lambda _{Q,h}^2\Lambda_Q^{d-3}\,|\omega |^{(d-1)/2}, & | \omega |>x_{\bf{k}} \\ 
\Lambda_{Q,h}^2\Lambda_Q^2\,\omega^2/(\delta{\bf k})^{5-d},  & |\omega |<x_{\bf{k}}.
\end{array}
\right. 
\label{a}
\end{equation}

The extension of the hot spots in momentum space for given energy $\omega $
can be read off (after reinstalling conventional units) as $(\delta k_{\bot
}/k_{F})\approx (\omega /\epsilon _{F})^{1/2}\Lambda _{Q}\cos \alpha $\ and $%
(\delta k_{\Vert }/k_{F})\approx (\omega /\epsilon _{F})^{1/2}\Lambda
_{Q}\sin \alpha $\ for deviations $\delta \mathbf{k}$ normal and parallel to
the Fermi surface, respectively. Here $\alpha $ is the angle subtended by $%
\mathbf{v}_{\mathbf{k}_{h}}$ and $\mathbf{v}_{\mathbf{k}_{h}+\mathbf{Q}}$,
which is expected to be of order $\pi /2$. We denoted the vertex
functions at the hot spots by $\Lambda _{Q,h}$, while $\Lambda _{Q}$
refers to the cold part of the Fermi surface.

From these results, we can find the real part of $\Sigma$ by
Kramers-Kronig transform and finally the quasiparticle $Z$-factor as
follows: 
\begin{equation}
{\rm Re}\Sigma^{(a)}({\bf k}, \omega) = \left(\int_{x_{\bf k}}^{\omega_0} +\int_0^{x_{\bf k}}\right)d\omega'\frac{\omega {\rm Im}\Sigma ^{(a)}(\mathbf{k},\omega ^{\prime })}{\omega
^{\prime 2}-\omega ^{2}}=\Sigma _h+\Sigma _l
\end{equation}
Carrying out the calculation for a wavevector ${\bf k}$ on a cold part of the Fermi surface, one finds that the self energy $\Sigma^{(a)}({\bf k},\omega)$ has no singular behavior in $\omega$. This may already be seen from the above expression for ${\rm Im}\Sigma$. Of course, it is expected that  singular behavior due to single spin fluctuation  scattering will only occur at the hot spots. Therefore, we may now take $x_{\bf k}  =(\delta\mathbf{k} /\Lambda _Q)^2\approx 0$ and find. 
\begin{eqnarray} 
\Sigma _h  &\approx -\omega\Lambda_{Q,h}^2\Lambda_Q^{d-3}\left( \int^{\omega_0}_0 d\omega'
\frac{\omega'^{(d-1)/2}}{\omega'^2-\omega^2} \approx \int^{\omega_0}_\omega \omega'^{(d-5)/2}d\omega' \right)\nonumber \\
&= -\omega\Lambda_{Q,h}^2\Lambda_Q^{d-3}|\omega|^{(d-3)/2}
\\
\Sigma_l &\approx 0
\end{eqnarray}
The contribution to the quasiparticle weight factor from spin fluctuations
is obtained as 
\begin{eqnarray}
Z_{S}^{-1}(\mathbf{k}_{h},\omega )& =1-\frac{\partial \mathrm{Re}\Sigma
^{(a)}(\mathbf{k}_{h},\omega )}{\partial \omega }  \nonumber \\
& =%
\left\{
\begin{array}{cc}
1-c_{Z,3}\Lambda _{Q,h}^{2}\ln |\omega |, & d=3 \\ 
1+c_{Z,2}\Lambda _{Q,h}^{2}\Lambda _{Q}^{-1}|\omega |^{-1/2}, & d=2.%
\end{array}
\right.
\label{z}
\end{eqnarray}

Here, the $c_Z$s  are constants of order unity. To be precise, the vertex $\Lambda _{Q,h}$ in this analysis connects a hot
spot $\mathbf{k}_{h}$ to $\mathbf{k}_{h}+\mathbf{Q+q}$ on the cold part of
the Fermi surface. 
This will be taken into account below. 

Inspection of Eq.~(\ref{Sigma_k})\ shows that the momentum dependence of 
}$\Sigma ^{(a)}(\mathbf{k},\omega )$ also becomes critical. However, it will  be shown later (Sec.~IV) that altogether, the critical renormalizations are so strong in the hot regions that a quasiparticle description is no longer valid there.

So far, we have assumed that the spatial dimension of fermions
and bosons is the same. However, it frequently happens that the  wavevector 
 of the spin fluctuations is quasi-two-dimensional, \textit{e.g.} located in the $xy$
plane, while the Fermi
surface is  that of a three-dimensional system.  In this case the manifold of hot momenta is a non-zero fraction of
the Fermi surface. For  momenta ${\bf k}$ in the manifold of hot momenta, in addition to the integrations above for $d=2$, there is the integral over the component $q_{z}$ along the
third dimension:
\begin{equation}
\int dq_{z}\mathrm{Im}G(\mathbf{k}_h\mathbf{+q},\omega +\nu )\propto \frac{1%
}{v_{F}}
\end{equation}
It follows that 
\begin{eqnarray}
\mathrm{Im}\Sigma ^{(a)}(\mathbf{k}_{h},\omega ) &\propto &u^{2}\Lambda
_{Q}^{2}N_{0}\xi _{0}^{-4}v_{F}^{-1}\int_{0}^{\omega }d\nu \frac{\nu }{%
\epsilon _{F}}\Lambda _{Q}^{2}\int_{q_{l}}\frac{dq}{q^{3}}  \nonumber \\
&\propto &\Lambda _{Q,h}^{2}|\omega |
\end{eqnarray}%
which is identical with the above result, Eq.~(\ref{z}) at $d=3$.

\subsubsection{Two-loop approximation}

\begin{figure}
{\normalsize \centering
\includegraphics[width=.65\textwidth, ]{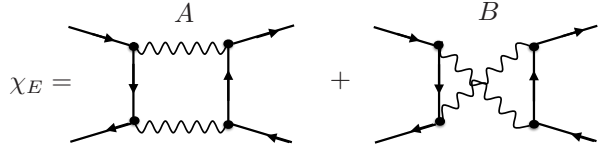} 
\vskip -2.5cm }
\caption{Structure of the energy fluctuation. The spin fluctuations $\protect%
\chi$ (wavy lines) carry momentum $\approx Q.$}
\label{chiE}
\end{figure}

 A contribution to the electron self energy that is critical
everywhere on the Fermi surface is generated by the exchange of \textit{two}
spin fluctuations, with nearly vanishing total momentum. Consider an
exchange energy density determined by $J {\vec S}_i\cdot{\vec S}_j$ Its
correlator has the form $K_E \sim \langle {(\vec S}_i\cdot{\vec S}_j)( {\vec
S}_k\cdot{\vec S}_l)\rangle$, which contains two spin fluctuation
propagators in its ``disconnected part":
\begin{equation*}
K_E\sim\langle S_i^+S_l^-\rangle\langle S_k^+S_j^-\rangle. 
\end{equation*}
This motivates our definition of an ``energy fluctuation" propagator $\chi _{E}$, built from two spin fluctuations, see Fig.~\ref{chiE} as 
\begin{eqnarray}
\mathrm{Im}\chi _{E}(k,k^{\prime };\mathbf{q},\nu )=& \sum_{\mathbf{q}%
_{1},\nu _{1}}G_{k+q_{1}}G_{k^{\prime }+q_{1}-q}\mathrm{Im}\chi (\mathbf{q}%
_{1},\nu _{1})  \nonumber \\
 & \times \mathrm{Im}\chi (\mathbf{q}_{1}-\mathbf{q},\nu _{1}-\nu )[b(\nu
_{1}-\nu )-b(\nu _{1})],  \label{b}
\end{eqnarray}%
where $b(\nu )$ is the Bose function and $k, k^{\prime },q$ denote
``four-momenta", \textit{e.g.} $(\mathbf{k},\omega).$ 
Assuming the external four momenta $k,k^{\prime }$ to be
``on-shell", \textit{i.e.} $(k\approx k_F,\omega\approx 0)$ \ and $q$
small,\ the Green's functions $G_{k+q_{1}},G_{k^{\prime }+q_{1}-q}$ are
off-shell for most values of the momenta $\mathbf{k,k}^{\prime },\mathbf{q}%
_{1}$\ (the cold part of the Fermi surface) and each may be replaced by $%
1/\epsilon _{F}$; this removes the dependence on $k,k^{\prime }$ from $\chi _{E}$.%
There are actually two relevant diagrams here as shown in Fig.~\ref{chiE}: diagram $(A)$ with parallel
and diagram $(B)$ with crossed $\chi $-lines.\ In the approximation of
replacing the intermediate $G^{\prime }s$ by $1/\epsilon _{F}$, the two
diagrams have identical momentum and frequency dependence, but differ in
spin dependence. Denoting the spins of the incoming and outgoing
particle-hole pairs by $(\alpha ,\beta )$ and $(\alpha ^{\prime },\beta
^{\prime })$ one finds  $\chi _{E;\alpha \beta ,\alpha ^{\prime }\beta ^{\prime
}}^{A,B}(k,k^{\prime };\mathbf{q},\nu )\propto \mp \lbrack 
3\delta _{\alpha \beta }\delta _{\alpha ^{\prime }\beta ^{\prime }} 
\pm 2\bf{\sigma }_{\alpha \beta }\cdot \bf{\sigma }_{\beta^{\prime }\alpha ^{\prime}}
]
$  Here the overall sign is different owing to an additional fermion loop
in diagram $A$. The sum of the two contributions gives a pure spin-spin dependence
$\chi _{E;\alpha \beta,\alpha ^{\prime }\beta ^{\prime}}\propto \bf{\sigma }_{\alpha
\beta }\cdot \bf{\sigma }_{\beta ^{\prime}\alpha ^{\prime}}$. 

Performing the momentum integration in Eq.~(\ref{b}) by Fourier
transform, one finds 
\begin{eqnarray}
\mathrm{Im}\chi _{E}(\mathbf{q},\nu )\approx & N_{0}^{3}\Lambda _{Q}^{2d-3}%
\big(\frac{|\nu|}{\epsilon _{F}}\big)^{d-1/2} \frac{1}{\left[ r + q^2
\xi_0^2+|\nu|\Lambda_Q^2/\epsilon_F \right]^{(d+1)/2}} ,  
\label{XE}
\end{eqnarray}%
One factor of $N_0$ comes from the momentum integration, the other two from the two spin fluctuation propagators.

The energy-fluctuation propagator $\chi _{E}$ may be considered as an
exchange boson that generates an effective interaction $V_{E}(\mathbf{q},\nu
)$ between quasiparticles. Since $\chi _{E}$ couples to a particle-hole (``p-h") pair, $V_{E}$ is screened by p-h polarization effects, represented by
the polarization function $\Pi(\mathbf{q},\nu)$, so that 
\begin{equation}
V_{E}(\mathbf{q},\nu )=\frac{u_E^{2}\left[\Lambda_E\cdot \chi _E(\mathbf{q},\nu)\right]}{%
1+ u_{E}^{2}\Pi (\mathbf{q},\nu )\left[\Lambda_E\cdot \chi _E(\mathbf{q},\nu)\right]}
\label{V_E}
\end{equation}
Here $\Lambda_E =\Lambda _{v}^{2}\Lambda _{Q}^{4}$ is an overall vertex factor arising from
vertex functions of two types: first, there arises a vertex function $%
\Lambda _{Q}$ at each end of a spin fluctuation propagator, on the cold part
of the Fermi surface. As already mentioned, it diverges as $Z^{-1}$ (Ref.~\cite{WA16} and below), second, at the
ends of the composite propagator $\chi _{E}$ a vertex function $\Lambda
_{v}$ arises. Both vertex corrections are generated only by irreducible
diagrams (the reducible parts are either incorporated in $\chi $, or
explicitly summed, as expressed in Eq.~(\ref{V_E}). The Ward identity based
on particle number conservation, gives $\lim_{\nu \rightarrow 0}\Lambda _{v}(%
\mathbf{q}=0,\nu )\propto Z^{-1}$, but see the discussion below Eq.~(\ref{Lh}).
Finally, $u_{E}\propto u^{2}\propto
N_{0}^{-2}$ is the bare interaction vertex connecting a p-h pair and an
energy fluctuation. We shall see below that $[\Lambda_E \cdot\chi _{E}(\mathbf{q},\nu
)]$ also diverges in the limit ${\bf q}=0,\nu \rightarrow 0$ as $Z^{-1}$. This might
lead to the expectation that the screening expressed in Eq.~(\ref{V_E}) will
remove the singular behavior of $V_{E}$. This is not the case, since in the
denominator of Eq.~(\ref{V_E}) the divergence of $[\Lambda_E\cdot \chi _{E}]$ is
removed by the vanishing of the quasiparticle polarization bubble $\Pi
\propto Z$. As introduced here, $\Pi $ is the bubble diagram without vertex
corrections, as those are included in $\Lambda_E $. The screening thus amounts
to a renormalization of $V_{E}$ by a factor of order unity, which we neglect
in the following, where we analyze quasiparticle scattering off energy fluctuations $\chi_E$.
We note in passing that the effective interaction of Eq.~(\ref{V_E}) gives rise to quantum corrections to the conductivity of disordered metals near an antiferromagnetic transition.\cite{Weiss16}

In one-loop approximation in terms of the quasi-boson propagator $%
\chi _{E}$, the imaginary part of the self energy in the cold regions is
given by 
\begin{eqnarray}
\mathrm{Im}\Sigma _{c}^{(b)}(\mathbf{k},\omega )& \propto u_{E}^{2}\Lambda
_{v}^{2}\Lambda _{Q}^{4}\int \frac{d\nu }{2\pi }\int \frac{d^{d}q}{(2\pi
)^{d}}\mathrm{Im}G(\mathbf{k+q},\omega +\nu )  \nonumber \\
& \times \mathrm{Im}\chi _{E}(\mathbf{q},\nu )[b(\nu )+f(\omega +\nu )].
\label{2loopSE}
\end{eqnarray}%
As in the evaluation of the one-loop self energy contribution of spin fluctuations, Eq.~(\ref{1}), the $\nu $%
-integration at low $T$ is confined to the interval $[0,\omega ]$ and $%
\mathrm{Im}G(\mathbf{k+q,}\omega +\nu )=\pi Z\delta (\omega +\nu -\epsilon _{%
\mathbf{k}}^{\ast }-v_{\mathbf{k}}^{\ast }q\cos \theta ),$ where $\theta $
is the angle enclosed by $(\mathbf{k,q)}$. The angular integral is
 \begin{equation}
\int d\cos \theta\, \mathrm{Im}G(\mathbf{k+q},\omega +\nu )\propto \frac{Z}{v_{%
\mathbf{k}}^{\ast }q}\propto \frac{1}{v_{F}q}.  \nonumber
\end{equation}%
${\rm Im}\chi_E$ is given in Eq.~(\ref{XE}). The $q$-integration may be done approximately by recognizing that the
integral is again controlled by the lower cutoff $q_{l}=\xi _{0}^{-1}(\nu
\Lambda _{Q}^{2}/\epsilon _{F})^{1/2}$, now determined by the denominator of Eq.~(\ref{XE}):
\begin{eqnarray}
\mathrm{Im}\Sigma _{c}^{(b)}(\mathbf{k},\omega )& \propto
u_E^2 N_{0}^{3}\Lambda _v^{2}\Lambda _{Q}^{2d+1}\int_{0}^{\omega }d\nu
\nu^{d-1/2}\,
q_{l}^{-2}(\nu)  \nonumber \\
& \approx \Lambda_v^2\Lambda _{Q}^{2d-1}|\omega |^{d-1/2}.
\end{eqnarray}%
The real part of $\Sigma $ follows by analyticity as 
\begin{equation}
\mathrm{Re}\Sigma ^{(b)}(\mathbf{k},\omega )\propto -\mathrm{sign}%
(\omega)\Lambda_v^2\Lambda _{Q}^{2d-1}|\omega |^{d-1/2}
\end{equation}%
and the contribution of energy fluctuations to the $Z$-factor at cold regions is thus obtained as 
\begin{equation}
Z_{c}^{-1}=1+b_{Z}\Lambda_v^2\Lambda _{Q}^{2d-1}|\omega |^{d-3/2},
\end{equation}%
where $b_{Z}$ is a constant of order unity. We recall that the exchange of
a single spin fluctuation does not lead to a scaling contribution to $Z^{-1}$
on the cold parts of the Fermi surface. If  ${\bf k}$ were to be at a hot region of the Fermi surface, these results still apply, with $Z_c$ replaced by $Z_h$. In the following, $Z_c$ is usually replaced by simply $Z$, as the cold regions are the majority of the Fermi surface.

Higher order contributions are likely to be irrelevant. In Appendix A we
show that two-loop contributions to the self energy in energy fluctuation language (four-loop
in the spin fluctuation framework) are smaller than the one-loop result $%
\propto \omega ^{1-\eta }$ by a factor $\omega ^{1-3\eta}\rightarrow 0$ as $\omega \rightarrow 0$ for dimensions $3/2<d<9/2$.  We introduced the exponent $\eta$ in Sec.~II for the frequency dependence of the $Z$-factor. It will be determined self-consistently as $\eta = (2d-3)/4d$  in Sec.~ IV. 
\subsubsection{Total self energy}
We now collect the results for the self energy.  On the cold parts of the
Fermi surface we only have a contribution from the scattering off energy
fluctuations 
\begin{eqnarray}
Z^{-1}_c &=&1+Y_c \\ \label{Zc}
Y_c &=&b_{Z}\Lambda_v^2\Lambda _{Q}^{2d-1}|\omega |^{d-3/2},
\label{Yc}
\end{eqnarray}%
whereas at the hot parts both energy and spin fluctuation scatterings
contribute 
\begin{eqnarray}
Z_{h}^{-1} &=&1+Y_{h}+X_h \\ \label{Zh}
Y_{h} &=&b_Z\Lambda_{v,h}^2 \Lambda _{Q}^{2d-1}|\omega |^{d-3/2}\label{Yh} \\
X_h &=&c_Z\Lambda _{Q,h}^2\Lambda_Q^{d-3}|\omega |^{(d-3)/2}  \label{Xh}
\end{eqnarray}%
In the latter, we keep in mind that $\lim_{\epsilon \rightarrow 0}|\omega |^{\epsilon
}\rightarrow \ln (1/|\omega |)$. 

\subsection{ Vertex function $\Lambda _{Q}$}
\begin{figure}[h]
 {\centering
\hspace{1.2in} 
\includegraphics[width=.5\textwidth,]
{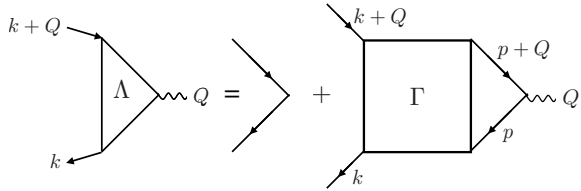}}
\vspace{-.6in}\caption{Structure of the vertex function $\Lambda_Q$. The spin fluctuation
 (wavy line) carries momentum $\approx Q$. $\Gamma$ is the ph-irreducible four-point function}
\label{V}
\end{figure}

To complete the self-consistent program for the determination of the $Z$ factors, we need expressions for the vertex functions.
The vertex $\Lambda _{Q}$ is obtained as the sum of
all p-h-irreducible three-point diagrams at non-zero momentum $\Lambda _{Q}(%
\mathbf{k},\omega ;\mathbf{q,}\nu )$, $\mathbf{q\approx Q}$,\ connecting a
particle-hole pair with momenta $(\mathbf{k+q},\omega +\nu )$ and $(\mathbf{k}%
,\omega )$ to a spin density fluctuation of momentum $(\mathbf{q},\nu )$.  See Fig.~\ref{V}. It
is important to realize that the vertex correction involves only
p-h-irreducible diagrams. The reducible diagrams are already incorporated in
the spin fluctuation propagator. As shown in Appendix B, the vertex
corrections  on the cold part of the Fermi surface arising from single spin fluctuation or energy fluctuation
exchange are small, in that they invariably include the small factor $|\omega |/\epsilon _{F}.$

\begin{figure}[b]
{\centering
\hspace{1.2in} 
\includegraphics[width=.28\textwidth, ]
{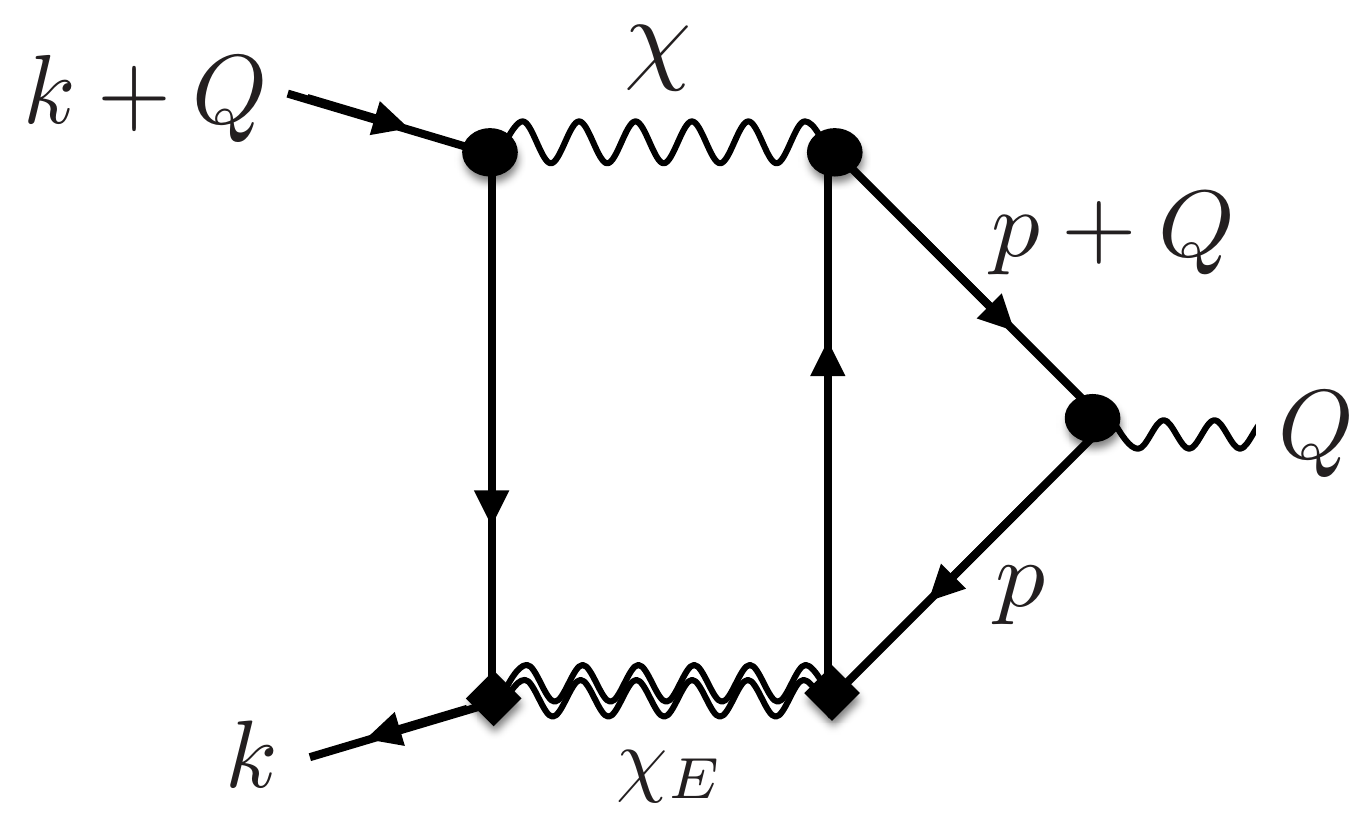}}
\caption{Two-loop diagram for  $\Lambda_Q$. The black dots ($\Lambda_Q$) and diamonds ($\Lambda_Q^2 \Lambda_v)$ signify  the appropriate vertex functions [see Eq.~(\ref{2loop})].}
\label{V2}
\end{figure}

This small factor (at $d=3$) is avoided in those diagrams in which the external
and internal momenta,, $k$ and $p$ as in Fig.~\ref{V}, are decoupled. The first such diagram (two-loop contribution) has
three $\chi ^{\prime }s$, two of them coupled to $\chi _{E}$ (see Fig.~\ref{V2}) 
\begin{equation}
\Lambda _{Q}(k;Q)=u^{6}\Lambda _{Q}^{6}\Lambda _{v}^{2}\int
(dq)G(k-q)T(q;Q)\chi (Q+q)\chi _{E}(q)
\label{2loop}
\end{equation}%
where $u\approx N_{0}^{-1}$ is the interaction vertex (we have altogether
six endpoints of the spin fluctuation propagator, with one interaction
vertex $u$ each).\ We define the quantity (triangle loop on the right side of Fig.~\ref{V2})
\begin{equation}
T(q;Q)=\int (dp)G(p-q)G(p+Q)G(p)
\end{equation}%
and approximate $T(q;Q)\approx T(0;0)\approx N_{0}/\epsilon_F$. We now use 
\begin{equation}
\mathrm{Im}\chi (\mathbf{q+Q},\nu )=\frac{N_{0}|\nu |\Lambda _{Q}^{2}}{%
(q^{2}+r)^{2}+(|\nu |\Lambda _{Q}^{2})^{2}}  \nonumber
\end{equation}%
where $\nu $\ and $q$ are dimensionless (frequencies and wave vectors in
units of $\epsilon _{F}$ and $k_{F}$\ ), and from Eq.~(\ref{XE}),
 \begin{equation}
\mathrm{Im}\chi _{E}(\mathbf{q},\nu )=\frac{N_{0}^3\Lambda
_{Q}^{2d-3}(|\nu |)^{d-1/2}}{[q^{2}+r+|\nu |\Lambda _{Q}^{2}]^{(d+1)/2}} 
\nonumber
\end{equation}%
The imaginary part of the
vertex function in the critical regime ($r=0$) is then approximately given
by\ 
\begin{eqnarray}
\mathrm{Im}\Lambda _{Q}(k;Q)& \propto \Lambda _{Q}^{6}\Lambda
_{v}^{2}\int_{0}^{\omega }d\nu \int dqq^{d-2}\frac{\Lambda _{Q}^{2d-1}(|\nu
|)^{d+1/2}}{[q^{2}+r+|\nu |\Lambda _{Q}^{2}]^{\frac{d+5}{2}}}  \nonumber \\
& \propto \Lambda _{v}^{2}\Lambda _{Q}^{2d-5}\int_{0}^{\omega }d\nu (|\nu
|)^{d+1/2}\int_{\Lambda _{Q}\sqrt{|\nu |}}\frac{dq}{q^{7}}  \nonumber \\
& \propto \Lambda _{v}^{2}\Lambda _{Q}^{2d-1}|\omega |^{d-3/2}.\label{IL}  \nonumber
\end{eqnarray}%
on the cold part of the Fermi surface (external particle and hole are both ``cold"). Here a factor of $1/q$ was generated by the angular integration of the Green's
function, as in Eq.~(\ref{theta}). A more complete account of this derivation msay be found in Ref.~\cite{WA16}.
The real part of $\Lambda
_{Q}(k;Q)$ is obtained by Kramers-Kronig transform. We add the bare
value $\Lambda _{Q}=1$, and find, on the cold part of the Fermi surface 
\begin{equation}
{\rm Re}\Lambda _{Q}(k;Q)=1+c_{\Lambda}\Lambda _{v}^{2}\Lambda
_{Q}^{2d-1}|\omega |^{d-3/2}+...    
\label {Lc}
\end{equation}%

In the hot region we have
(considering that only one of the external momenta will be at the hot spot,
as discussed at the end of section III), 
\begin{equation}
\mathrm{Im}\Lambda _{Q,h}\propto \Lambda _{v}\Lambda _{v,h}\Lambda
_{Q}^{2d-1}|\omega |^{d-3/2}  \nonumber
\end{equation}%
Here we take into account that the vertex $\Lambda _{v,h}$  connected to the incoming hot quasiparticle or quasihole is generally different from the vertex  $\Lambda _{v}$ connected to the outgoing cold quasiparticles.
Then, at the hot spots 
\begin{equation}
{\rm Re}\Lambda _{Q,h}(k;Q)=1+c_{\Lambda}\Lambda _{v,h}\Lambda
_{v}\Lambda _{Q}^{2d-1}|\omega |^{d-3/2}+....  
\label{Lh}
\end{equation}%
We see that $\Lambda _{Q}$ depends on the position at the Fermi surface
through the external vertex function factors $\Lambda _{v}(\mathbf{k};%
\mathbf{q})$. In fact, even if we consider the self energy at a hot spot $%
\mathbf{k}_{h}$, only one of the incoming momenta, say $\mathbf{k}$, is hot,
while the other one $\mathbf{k}+\mathbf{q}$ (where $\mathbf{q}$ is the
momentum transferred to the energy fluctuation) is typically cold. Ward
identities for vertex functions at non-zero $\mathbf{q}$ are discussed in
Ref.~\cite{WA16}. There it is shown that the full vertex (including
reducible parts) is related to the $Z$-factors of the particle and hole
lines by $\Lambda _{v}(\mathbf{k};\mathbf{q})\approx \frac{1}{2}(Z_{\mathbf{k%
}}^{-1}+Z_{\mathbf{k}+\mathbf{q}}^{-1})$. This property  is expected to carry over to the
irreducible vertex as well. We therefore conclude that $\Lambda _{v}(\mathbf{%
k}_{h};\mathbf{q})\approx \Lambda _{Q,h}\approx \frac{1}{2}%
(Z_{h}^{-1}+Z^{-1})$ whereas for cold quasiparticles $\Lambda _{v}(\mathbf{k}%
;\mathbf{q})\approx Z^{-1}$. 

\section{ Self-consistent determination of self energy and
vertex functions}\label{sc}

 The expressions for the self energy and the vertex functions
derived above form a set of self-consistent equations for $Z^{-1},Z_{h}^{-1},
$ and $\Lambda _{Q},\Lambda _{Q,h}$. We now discuss the solution of these
equations in the limiting cases of weak and strong coupling.
 In the case of strong coupling we assume $Z^{-1}\gg 1$ and $%
\Lambda _{Q}\gg 1 $ . We first consider the cold regions. Combining
Eqs.~(\ref
{Yc},\ref{Lc}), we find $Z^{-1}=\Lambda _{Q}$ and hence\ 
\begin{equation}
Z^{-1}\propto |\omega |^{-(d-3/2)/2d} = |\omega |^{-\eta}  
\label{eta}
\end{equation}%
which is the power law  already found in Ref.~\cite{asw} and from which $\eta  = (2d
-3)/4d$ as quoted earlier.

The weak coupling behavior is found as 
\begin{eqnarray}
Z^{-1} &=&1+b_Z|\omega |^{d-3/2}  \nonumber \\
\Lambda _{Q} &=&1+c_{\Lambda}|\omega |^{d-3/2}
\end{eqnarray}

At the hot spots we have two contributions to the self-energy,
and $Z_{h}^{-1}=1+X_{h}+Y_{h}$. By examining the powers of $\omega$ and $\Lambda$ in Eqs.~(\ref{Yh},\ref{Xh}), we determine that $X_{h}\gg Y_{h}$. The
external momenta of the self-energy are by assumption at the hot spots,
while the internal momenta of the Green's function (in $\Sigma= \int \chi G$\,) are not, if we take
typical values of the momentum $\mathbf{q}$.  Then
the vertex function at the hot spots Eq.~(\ref{Lh}) is obtained as
\begin{equation}
\Lambda _{Q,h}\propto (Z_{h}^{-1}+Z^{-1}) \nonumber
\end{equation}
If we substitute this $\Lambda _{Q,h}$
for the vertex function in Eq.~(\ref{Xh}), we get
 \begin{equation}
Z_{h}^{-1} = 1 + c_Z' (Z_h^{-1}+Z^{-1})^2Z^{-d+3}|\omega |^{(d-3)/2}
\label{Z+h}
\end{equation}%

At weak coupling, we find
\begin{equation}
Z_h^{-1} = 1+ c_Z'|\omega |^{(d-3)/2} 
\end{equation}
The behavior at strong coupling cannot be read off directly from the
self-consistent equations. Inspection of Eq.~(\ref{Xh}) shows that $%
Z_{h}^{-1}>Z^{-d+1}|\omega |^{(d-3)/2}>|\omega |^{-(\frac{1}{4}+\frac{3}{4d}%
)}$. The exponent $\eta=(\frac{1}{4}+\frac{3}{4d})>\frac{1}{2}$ implies that
quasiparticles are no longer well-defined at the hot spots in the strong
coupling regime. As explained in Sec.~II, $\eta$ less than about 0.36 is required.
Within our approach, which assumes the existence of
well-defined quasiparticles, we are not equipped to describe the properties
of incoherent fermionic excitations at the hot spots. As long as the
extension of the hot spots is sufficiently small, their contribution to the
thermodynamic and transport properties may be expected to be small, too. We
will therefore discard the effect of hot quasiparticles in the following.  

\section{Renormalization group flow}

The analysis of the previous paragraphs reveals that for sufficiently
strong interactions, the low-energy properties of a system near a metallic
quantum critical point is governed by new scaling behavior. As we discussed in the previous section, the limiting cases of strong and weak coupling may exhibit different power-law (scaling) exponents  as $\omega \to 0$. It is
a natural question to ask whether this behavior can be expressed
in terms of renormalization group (``RG") equations involving corresponding $\beta$-functions that determine the
renormalization group flow of the self energies of hot and cold quasiparticles. We shall express the RG flow in terms of dimensionless ``coupling
constants", $\alpha_{h}$ and $\alpha_{c}$ that are related to the self energies  for the hot and cold regions
 \begin{equation}
\frac{d\alpha_{l}}{d\log\left(\epsilon_{F}/|\omega|\right)}=\beta_{l}\left(\alpha_{c},\alpha_{h}\right).
\end{equation}
with $l\in\left\{h,c\right\} $.  As suggested by us previously in Ref.~\cite{asw},  a natural definition of the dimensionless
coupling constants is in terms of the $Z$-factor:  $\alpha_l=Z_l^{-1} -1$.

To consider the associated renormalization group flow for cold
particles we use $\alpha _{c}$ with the results of Sec.~III for the self
energy. The scaling quantity is $\alpha _{c}(t)\ $as a function of $t=\ln
(\epsilon _{F}/|\omega |)$. It is given by [see Eq.~(\ref
{Yc})]
\begin{equation}
\alpha _{c}=b_{Z}(1+\alpha _{c})^{2d+1}e^{-(d-3/2)t}
\end{equation}%
Taking the derivative of $\alpha _{c}$ with respect to $t$, we get 
\begin{equation}
\frac{d\alpha _{c}}{dt}=-(d-3/2)\alpha _{c}+(2d+1)\frac{\alpha _{c}}{%
1+\alpha _{c}}\frac{d\alpha _{c}}{dt}
\end{equation}%
Solving for $d\alpha _{c}/dt$, we find the flow equation 
\begin{equation}
\frac{d\alpha _{c}}{dt}=\beta _{c}(\alpha _{c})=\frac{-(d-3/2)\alpha _{c}(1+\alpha _{c})}{1-2d\alpha
_{c}}.
\end{equation}%
The $\beta $-function for cold quasiparticles only depends
on the coupling constant $\alpha _{c}$ itself.

For weak coupling,  $\alpha_{c}\ll1$, 
\begin{equation}
\beta_{c}\left(\alpha_{c}\ll1\right)\approx-\left(d-\frac{3}{2}\right)\alpha_{c},
\end{equation}
{\it i.e.} for $d>\frac{3}{2}$ the interaction between cold carriers is
irrelevant and the flow of $\alpha_{c}$ is directed towards the stable
fixed point at $\alpha_{c}=0$. In the neighborhood
of this fixed point one finds 
\begin{equation}
\alpha_{c}\propto \left| \omega \right| ^{d-3/2}, \, \, \alpha_{c}\ll1/(2d),
\end{equation}
This corresponds to subleading corrections to Fermi liquid scaling,
in full aggreement with the results found by Hartnoll et al.\cite{Hartnoll11}
for $d=2$. For weak coupling, deviations from Fermi-liquid physics
are limited to hot portions of the Fermi surface (see below). For
$d>2$ it is furthermore possible to integrate out the fermionic degrees of freedom
(see Ref.~\cite{Abanov00}) and one recovers the physics described by the
Hertz-Millis-Moriya theory.
 
\begin{figure}[h]
 {
\vspace{-1.1in} \includegraphics[width=0.9\textwidth,]
{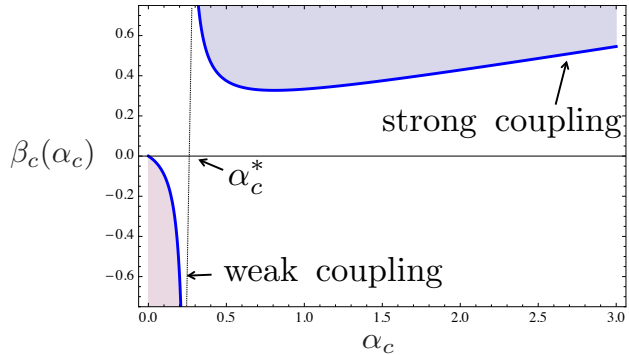}}
\vspace{-1.3in}
\caption{ $\beta$-function of the renormalization group flow of cold quasiparticles for $d=2$. While the flow is towards weak coupling if $\alpha_c < \alpha_c^*$, the system flows to infinite coupling with power law behavior for $\alpha_c > \alpha_c^*$ . The two regimes, corresponding to weak and strong coupling, are separated by a pole of the $\beta$-function. }
\label{VI}
\end{figure}

For large values of the coupling constant
we obtain 
\begin{equation}
\beta_{c}\left(\alpha_{c}\gg1\right)\approx\frac{d-3/2}{2d}\alpha_{c}.
\end{equation}
Now, the system flows to strong coupling,  $\alpha_{c}\to\infty$;
it diverges according to a power law 
\begin{equation}
\alpha_{c}\propto|\omega|^{-\eta},\, \, \alpha_{c}\gg\alpha_{c}^{\ast}
\end{equation}
with same anomalous exponent as determined from the self consistency
argument discussed above: 
\begin{equation}
\eta=\frac{d-3/2}{2d}.
\end{equation}
The threshold value of the coupling constant that separates conventional
and critical Fermi liquid behavior is $\alpha_{c}^{*}=1/\left(2d\right)$,
where the $\beta$-function has a pole. It defines a separatrix that distinguishes the scaling in  the strong and weak coupling regimes. While a pole in the $\beta$-function
is rather unusual, it is not without precedent. A prominent example
is the instanton-based derivation of the $\beta$-function of supersymmetric
Yang-Mills theories in Ref.~\cite{Novikov83}. Whether the pole in $\beta\left(\alpha\right)$
of our theory is a robust phenomenon or will be smoothed is unclear
at this point. If smoothed, conventional and critical Fermi liquid
behavior would be separated by an unstable fixed point at $\alpha_{c}^{\ast}=1/2d$,
where $\beta$ changes sign. Thus, if one increases the value of the
coupling constant, right at the antiferromagnetic quantum critical
point, the system passes through yet another quantum critical point
that separates conventional and unconventional behavior. This is illustrated
in Fig.~\ref{VI}

\section{Scaling of dynamic and thermodynamic quantities}

In this section we discuss the observable quantities as calculated with our
theory in the strong coupling regime, making use of the results derived in
the earlier sections. We recall that the contribution from hot quasiparticles is expected to be small. While the
results below have been published before, for the most part the
presentation given here is somewhat different. We also review the comparison
of our results with experimental findings for two of the quantum critical
heavy fermion compounds, YbRh$_{2}$Si$_{2}$ (YRS) and CeCu$_{6-x}$Au$_{x}$
at $x=0.1$ (CCA), which we believe to be described well by the strong
coupling theory.

The spin fluctuations in YbRh$_{2}$Si$_{2}$, as studied by neutron
scattering \cite{Stock12} show two components: a
three-dimensional ferromagnetic component in a somewhat higher temperature
range, 0.3 K$<T<$10 K, that gives rise to non-Fermi liquid behavior and an
incommensurate antiferromagnetic component below $T\approx 0.3$ K, which is presumably
responsible for the quantum critical behavior observed at the field-tuned
QCP. The critical field is low, $H_c \approx 60$ mT and the AFM critical
temperature is also low, $T_N(H=0)\approx 70$ mK, well separated from the
Kondo scale of $T_K$ $\approx 20$ K estimated for this heavy fermion
compound. As the system is cooled down in the critical field, the
ferromagnetic fluctuations cause a logarithmically increasing effective mass
(or equivalently $1/Z$), as observed experimentally taking the system into
the strong coupling domain (as discussed at the beginnning of Sec.~IV), when the low temperature regime dominated by
antiferromagnetic fluctuations is reached.

For CeCu$_{6-x}$Au$_{x}$ the spin fluctuations have been studied in much
more detail (large crystals, no neutron absorbing elements).
Quasi-two-dimensional critical incommensurate antiferromagnetic fluctuation\cite{Schroeder00,Stockert10} are found and in addition, two-dimensional
ferromagnetic fluctuations.\cite{Rosch97} We remark that the hot regions of
the Fermi surface generated by quasi-two-dimensional antiferromagnetic
fluctuations in a three-dimensional metal are extended, i.e. cover a non-zero
region of the Fermi surface. Both of these fluctuations may be expected to
drive the system into the strong-coupling regime of a quasi-two-dimensional
critical antiferromagnetic system as we considered above.

In the following we report the comparison of the theoretical result for
each observable considered with available experimental data for the two
compounds, YRS and CCA. Such comparisons were illustrated with figures in our earlier papers on the subject; these are cited in the appropriate places.

\begin{figure}[h]
 \centering
\vspace{-.8in} \includegraphics[width=.89\textwidth, ]
{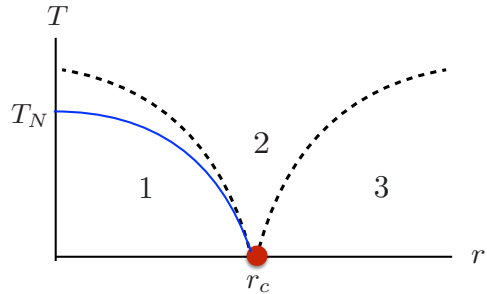} \vskip -3.7cm 
\caption{Sketch of example quantum critical  phase diagram. Temperature $T$ {\it vs} control parameter $r$.  $r_c$ marks the quantum critical point. Region 1: ordered regime, {\it e.g.} antiferromagnet, with N{\' e}el temperature $T_N$ at $r=0$. Region 2: quantum critical regime. Region 3: quantum disordered regime, often a paramagnetic Fermi liquid region. Dashed lines represent crossovers into the quantum critical regime. Blue line represents the thermodynamic N{\'e}el transition into the antiferromagnetic state.}
\label{qcp}
\end{figure}

In Fig.~\ref{qcp}, we give a sketch of a quantum critical phase diagram which identifies the various regions to which we refer below in our comparisons to data,

\subsection{Correlation length of bosonic fluctuations
and dynamical scaling}

We begin by discussing the correlation length $\xi $ of spin fluctuations,
which is identical to the one of energy fluctuations. Here and in the
following, we only consider the strong-coupling regime. It follows from 
Eq.~({\ref{X}) that $\xi =\xi _{0}r^{-1/2}$. The
renormalized control parameter $r$ is scale dependent; it depends on
temperature $T$ and frequency $\nu $, and its limiting value $\lim_{\{T,\nu
\}\rightarrow 0}r\equiv r(0)$. In the critical regime (region 2 of Fig.~\ref{qcp}), {\it i.e.} when we may put $%
r(0)=0$, the scaling may be read off  Eq.~({\ref{X}) as $\xi
(T,\omega )\propto \lbrack \max \{T,\omega \}]^{-1/z}.$ Here $z$ is the
dynamic critical exponent, obtained by determining the scaling relation $\nu
\propto q^{z}$ from the denominator of  Eq.~({\ref{X}). By using the scaling results of previous sections for $\Lambda_Q \sim Z^{-1}\propto |\omega|^{-\eta}$, one
finds
\begin{equation}
z=2/(1-2\eta )=4d/3.
\end{equation}%
It remains to determine the scaling of the correlation length with an
external field representing the control parameter. For definiteness we
consider a magnetic field tuned quantum phase transition in the following.
The divergence of the correlation length at $\{T,\omega \}\rightarrow 0$, as
a function of $H$, is by definition given as $\xi (H)\propto r_{0}^{-\nu }$,
where $r_{0}=|H/H_{c}-1|$. The correlation length exponent $\nu $ may be
determined as follows. We know that $r(0)$ must vanish at the QCP as a power
of $r_{0}$\ , which in general is not linear, for the following reason.\ A
uniform magnetic field $H$ couples to the system via the spin density of
conduction electrons. The corresponding coupling constant is critically
renormalized near the QCP, as represented by a three-legged vertex $\Lambda $%
, in the limit momentum $q\rightarrow 0$. This vertex is known to be equal
to $1/Z$, by virtue of the usual Ward identity. We thus have $r(0)\propto
r_{0}/Z$. The scaling of $Z$ with $|H-H_{c}|$ may be derived from the
scaling of $Z$ with $\omega $, $Z\propto \omega ^{\eta }$ , by noting that
the two scaling relations for $\xi $, inside and outside the critical cone,
have to match at the crossover, implying $\omega \propto r_{0}{}^{\nu z}$.
It follows that $r(0)\propto \xi ^{-2}(H)\propto r_{0}{}^{1-\eta \nu z}$,
from which one finds the correlation length exponent as
\begin{equation}
\nu =2/(2+z\eta )=3/(3+2d)
\end{equation}%
To summarize, the inverse correlation length (in units of $k_F$) may be expressed as
\begin{equation}
\xi ^{-1}(T,H)\propto T^{1/z}+c_{\xi }r_{0}{}^{\nu }
\end{equation}%
with $T$ in units of $\epsilon _{F}$ and $c_{\xi}$ a constant of order
unity. The above discussion applies as well to the case of tuning by
pressure, where $r_{0}=|p/p_{c}-1|$, or chemical composition, where $%
r_{0}=|x/x_{c}-1|$, in which cases the corresponding particle density vertex
is likewise critically renormalized by a factor $1/Z$.

In Ref.~\cite{asw}, we developed the scaling of the spin fluctuation spectral function; it shows the following general scaling
behavior: 
\begin{equation}
{\rm Im} \ \chi (\mathbf{q},\omega )\propto T^{-2/z}\Phi _{\chi }(\frac{\omega 
}{T},q\xi ,r_{0}T^{1/z\nu })
\end{equation}%
where the scaling function is given as \cite{asw}
\begin{equation}
\Phi _{\chi }(x,u,w)=\frac{x(x^{2}+a^{2})^{-\eta }}{(1+w^{2\nu
})[1+u^{2}]+[x(x^{2}+a^{2})^{-\eta }]^{2}}  \label{Xscale}
\end{equation}%
Here, the constant $a$ mediates the relation between energy scales $\omega$ and $T$.
{\it e.g.} $\omega^\gamma \to(\omega^2+a^2T^2)^{\gamma/2}$. In the critical regime, where we may put }$r_{0}=0$, the function $T^{2/z}\,%
{\rm Im}\chi (\mathbf{q},\omega )$ obeys $\omega /T$-scaling.

The dynamical scaling near the QCP of CeCu$_{6-x}$Au$_{x}$ at $x=0.1$\ has
been studied using inelastic neutron scattering.\cite%
{Schroeder00,Stockert10}} In this material, the antiferromagnetic spin
fluctuations appear to be quasi-two-dimensional. In the critical regime ($%
r_{0}=0$) the data obey $\omega /T$-scaling, of the form shown in 
Eq.~({\ref{Xscale})}, where the exponent $\eta $ has been
determined as $\eta _{ex}\approx 0.1-0.15$ , while our theory gives $\eta
_{th}=1/8=0.125$\ . A detailed comparison of experiment and theory, showing
very good agreement, is given in Ref.~\cite{asw}.

\subsection{Free energy}

The critical behavior of the free energy may be derived from the critical
behavior of the fermionic quasiparticles by expressing the free energy
density in terms of the fermionic self-energy

\begin{equation}
f(T,r_{0})=\frac{1}{2\pi }\int_{0}^{T}dT'N_{0}\int \frac{d\omega }{%
(T') ^2\cosh ^{2}(\omega/2T')}\int d\Omega _{\mathbf{k}%
}[\omega -\Sigma (\mathbf{k}_{F},\omega )]
\end{equation}%
where }$\int d\Omega _{\mathbf{k}}$ is a normalized angular
integral over the Fermi surface. It may be shown that the contribution of
the bosonic fluctuations to the free energy is subleading. Substituting the
results for $\Sigma $ obtained above, we find that in the strong-coupling regime, $f$ obeys the scaling 
\begin{equation}
f(T,r_{0})=\xi ^{-(2d+1)}\Phi _{f}(r_{0}\xi ^{1/\nu },T\xi ^{z})
\label{fscale}
\end{equation}%
This may be rewritten as 
\begin{equation}
f(T,r_{0})=T^{(2d+1)/z}\Psi (r_{0}{}^{z\nu }/T)
\end{equation}%
where the exponent of the prefactor may be expressed as $(2d+1)/z=2-\eta $
\begin{equation}
\Psi (x)\propto \lbrack 1+c_{f}x^{\eta }]^{-1}
\end{equation}%
where $c_{f}$ is a constant of order unity, and the second term in the
square brackets ensures that $f(T,H)\propto T^{2}$ in the Fermi liquid
regime, where $x\gg 1$. In  Eq.~({\ref{fscale}) the "correlation
volume" enters as $\xi ^{2d+1}$ rather than $\xi ^{d+z}$. This is a
consequence of the fact that the dominant critical degrees of freedom are
the fermionic ones. It may be shown that $\xi ^{2d+1}=\xi _{f}^{d_{f}+z_{f}}$%
, with the fermionic correlation length given by $\xi _{f}^{-1}\propto
T/Z(T)\propto T^{1-\eta }$, which grows much faster than the bosonic
counterpart $\xi ^{-1}\propto T^{1/z}$. The fermionic dimension $d_{f}=1$
and the fermionic dynamical exponent $z_{f}=1/(1-\eta )$. The fermionic
correlation length exponent $\nu _{f}=1$.\ In the literature on quantum
critical phenomena in metals the bosonic correlation length $\xi $ is
usually presented as the relevant length scale and we follow this usage.

Taking derivatives of the above free energy expressions, we may derive the
critical behavior of thermodynamic quantities. We start with the specific
heat coefficient $\gamma =C/T$ , found as the second derivative of $f$ with
respect to temperature:
\begin{equation}
C(T,r_{0})/T\propto r_{0}{}^{-(2d-3)/(2d+3)}+c_{\gamma }T^{-(2d-3)/4d}.
\end{equation}%
This implies that $C/T$  diverges in dimensions $d=3,2$ as a power law, both
when approaching the QCP at $r_{0}=0$ as $T\rightarrow 0$ or else at $T=0$
as $r_{0}\rightarrow 0$ from the Fermi liquid phase.

In the quantum critical regime, $r_{0}=0$, one thus finds for $3d$ and $2d$ fluctuations, $C/T\propto T^{-1/4}$ and $C/T\propto T^{-1/8}$, both
in good agreement with the data on YRS\cite{Oeschler08, qp1}
and on CCA,\cite{hvl96,asw} respectively. In the case of YRS, a classical contribution $%
C/T\propto 1/T$ was included; it may be traced to the excitation of
spin resonance bosons \cite{WA15} of energy lower than the lowest $T$ accessible in
experiment. Approaching the QCP from the Fermi liquid side our theory
predicts $C/T\propto r_{0}^{-1/3}$ for $3d$-fluctuations, in excellent
agreement with data on YRS.\cite{Custers03,Hartmann10}

Next we calculate the magnetization $M=-\partial f/\partial
H=-\partial f/\partial r_0 $ and find:
\begin{equation}
M(T,H)-M(0,H_{c})\propto -r_0^{2d/(2d+3)}-c_{M}T
\end{equation}%
which represents nonanalytic (cusp-like) behavior, as the QCP is approached.
In the critical regime the linear $T$-law amended by a Fermi liquid $T^{2}$%
-component has been shown to account well for the YRS-data \cite{Tokiwa09,pnas} and the CCA-data.\cite{hvl96,asw}

The magnetic susceptibility is given by a further derivative with respect to $r_0$:
\begin{equation}
\chi (T,H)-\chi (0,H_{c})\propto -r_{0}{}^{(2d-3)/(2d+3)}-c_{\chi
}T^{(2d-3)/4d},
\end{equation}%
which as a function of either $T$ or $H$ shows a downward cusp. In the
critical regime, and for $d=3$ we find the power law $T^{1/4}$, which,
when augmented by the Fermi liquid $T^{2}$ component, accounts very well for the
data on YRS. \cite{Gegenwart02,pnas}

A quantity of special interest is the Gr{\" u}neisen ratio, defined as the ratio
of entropy derivatives
\begin{equation}
\Gamma _{G}=-\frac{(\partial S/\partial X)_{T}}{T(\partial S/\partial T)_{X}}
\end{equation}%
where $X$ denotes a control parameter field. For
concreteness we consider the case, appropriate for YRS,  $X=H$, where $H$ is an applied magnetic field, in the following. Then,
the magnetic Gr{\" u}neisen ratio is
\begin{equation}
\Gamma _{G}=-\frac{\partial M/\partial T}{T(\partial S/\partial T)_H}= \left\{
\begin{array}{cc} -G_r/|H-H_c|, \;\;\;\; T\to 
0\\
c_\Gamma\,T^{-(2d+3)/4d},  \;\;\;\; H \to H_c.
\end{array} \right.
\end{equation}%
Here we find universal behavior in the Fermi-liquid regime, with the
coefficient $G_{r}=-\nu [z-(2d+3)/3]=-\nu _{f}(z_{f}-d_{f})$, in
agreement with the result following from a phenomenological analysis
that assumes hyperscaling.\cite{Zhu03} The magnetic Gr\"{u}neisen
ratio has been experimentally determined for YRS\cite{Tokiwa09}%
to follow a $T^{-0.7}$-law in the critical regime, and a coefficient $%
G_{r}\approx -0.3$ in the Fermi liquid regime. This compares well with the
theoretical results \cite{pnas} $T^{-3/4}$ and $G_{r}=-1/3$. In
CCA, $\Gamma _{G}$ appears to grow logarithmically with decreasing
temperature, and hence much slower than the predicted $T^{-7/8}$-law. The
coefficient $G_{r}$ has not  yet been determined.\cite{hvl}

\subsection{Transport and relaxation properties}

The basic inelastic relaxation rate is obtained from the imaginary part of
the self-energy. As discussed in Sec.~II, the quasiparticle relaxation rate is
defined as $\Gamma =Z \ {\rm Im}\Sigma $; it obeys the scaling law\cite{pnas}
\begin{equation}
\Gamma =\xi ^{-z}\Phi _{\Gamma }(r_{0}\xi ^{1/\nu },T\xi ^{z})
\end{equation}%
which may be reexpressed as
\begin{equation}
\Gamma (T,H)=T\Psi _{\Gamma }(r_{0}{}^{z\nu }/T)
\end{equation}%
with
\begin{equation}
\Psi _{\Gamma }(x)\propto \lbrack 1+c_{\Gamma }x]^{-1}
\end{equation}%

The critical behavior of the electrical resistivity $\rho (T,H)$ may be
obtained in the limit of low temperatures, where the contribution of
inelastic processes is small, so that the elastic scattering contribution $\rho(0,H)$ dominates. Then we may write
\begin{equation}
\rho (T,H)-\rho (0,H_{c})\propto \langle q^{2}\rangle \Gamma /Z
\end{equation}%
The factor  $\langle q^{2}\rangle $ accounts for vertex corrections, and is
approximately given by the typical momentum transfer in a scattering process
squared. Under the assumption made above that impurity scattering
dominates, the relevant inelastic processes are dressed by elastic
processes such that the typical momentum transfer is $k_F$, and is thus
noncritical. In this case we may put $\langle q^{2}\rangle \approx
c_{imp}k_F^{2}$, where $c_{imp}$ is the impurity concentration. We then
find
\begin{equation}
\rho (T,H)-\rho(0,H_c) \propto c_{imp}
 \left\{
\begin{array}{cc}
T^{1-\eta},
\;\;\;\;\;\;\;\;\;\;\;\;\;\;\;\;\;\; \;\;\;\;\;\;\;\;\; \;H\to H_c \\%
T^{2} |H-H_c|^{-(3z/2-1)\nu}, \;\;T\to 0.
\end{array}\right.
\end{equation}%
In the opposite case of clean samples (but still in the regime $\rho
(T,H)/\rho (0,H_c)-1 \ll 1$) the typical momentum transfers are
small, $\langle q^{2}\rangle \propto T/Z^2\propto T^{1-2\eta}$ in the
critical regime and $\propto T|H/H_{c}-1|^{-2(2d-3)/(2d+3)}$ in the Fermi
liquid regime. We find
\begin{equation}
\rho (T,H)-\rho (0,H_{c})\propto 
 \left\{
\begin{array}{cc}
T^{2-3\eta}, \;\;\;\;\;\;\;\;\;\;\;\;\; H \to H_c\\
T^3/|H-H_c |, \;\;\;T \to 0
\end{array}\right.
\end{equation}%
In the case of YRS ($3d$ spin fluctuations), we then expect $\rho (T,H)-\rho (0,H_c)\propto
c_{imp}T^{3/4},$ in good agreement with experiment.\cite%
{Gegenwart08,qp1} A comparison of the data in Refs.~\cite{Gegenwart08,qp1} with those of somewhat dirtier samples \cite{Taupin15} shows that the prefactor of the $T^{3/4}$-law indeed appears to
grow with impurity concentration. For
CCA ($d=2$), CeCu$_{6-x}$Au$_{x}$, for $x=x_c=0.1$, we find $c_{imp}T^{7/8}$ in reasonable agreement with the data of Ref.~\cite{hvl14}. These data were previously fit by a
linear $T$ law, which describes the data almost as well. AFM order  is found for $x\geq 0.15$ but  the $T_N$ can be tuned to zero by hydrostatic pressure ($p_c\approx 5$ kbar) or magnetic field($H_c\approx 0.4$ T).\cite{hvl01} 
It was found that for pressure tuning, the critical behavior of the resistivity follows an approximately linear $T$-law,
in agreement with the behavior found at the critical concentration $x=0.1$.
However, the $\rho$ data obtained by magnetic field tuning showed a $T^{3/2}$-law, as
is expected for a weak-coupling spin density wave QCP.  The contrast is instructive: it shows that  for CCA with $x=0.2$, both strong (pressure tuning) and weak (field tuning) coupling critical behavior may occur. We  may interpret this within our theory with the conjecture that  the magnetic
field of $H_{c}\approx 0.4T$ needed to reach the QCP suppresses the magnetic
fluctuations thought to be necessary to get into the strong coupling regime.
In this study,\cite{hvl01} it was also found that the
prefactor of the $T$-law appeared to increase with disorder.

The dynamical conductivity $\sigma (\omega +i0)$ may be represented as
\begin{equation}
\sigma (\omega +i0)=\frac{\omega _{p}^{2}}{4\pi }\frac{i}{\omega -M(\omega
+i0)}
\end{equation}%
where $\omega _{p}^{2}=4\pi e^{2}n/m$\ is the plasma frequency squared and $%
M(\omega +i0)$ is the relaxation kernel. $%
\sigma (\omega +i0)$ at not too high frequencies, $\omega \ll \epsilon _{F}$,
follows, using $\omega -M(\omega +i0)=\omega Z(\omega )+i/\tau (\omega )$ ,
where $Z(\omega )=1-{\rm Re}M(\omega )/\omega $\ and $1/\tau (\omega )=-%
{\rm Im}M(\omega )$ as
\begin{equation}
\frac{\omega _p^2}{4\pi }\,\frac{\tau(\omega)(1+i\omega\tau(\omega)Z(\omega)}{1+[\omega Z(\omega)\tau(\omega)]^2}
\end{equation}
We note that $Z(\omega )$ may be interpreted as inverse effective mass ratio 
$Z=m/m^{\ast }$.\ 

If we neglect vertex corrections we may relate $M$ to the self energy $%
\Sigma $ as
\begin{equation}
M(\omega +i0)=2\Sigma (\omega +i0)+i/\tau _{imp}  \nonumber
\end{equation}%
where $\tau _{imp}$ is the momentum relaxation time due to impurity
scattering.

Near a quantum critical point and in the critical regime (control parameter $%
r_{0}=0$) we expect the scaling behavior of the effective mass ratio

\begin{eqnarray}
Z(\omega ,T;r& =0)\propto T^{\eta }\Phi _{Z}(|\omega |/T, r=0), \\
\Phi _{Z}(x,0)& \propto x^{\eta }, \, \, x>>1  \nonumber \\
\Phi _{Z}(x,0)& =1, \, \, x<<1  \nonumber
\end{eqnarray}%
and of the relaxation rate
\begin{eqnarray}
\tau ^{-1}(\omega ,T;r& =0)=c_{in}T^{1-\eta }\Phi _{\tau }(|\omega |/T, r=0)+1/\tau _{imp}, \\
\Phi _{\tau }(x,0)& \propto x^{1-\eta },\, \, x>>1  \nonumber \\
\Phi _{\tau }(x,0)& =1, \, \, x<<1  \nonumber
\end{eqnarray}%
where $\eta $ is the critical exponent introduced in Sec.~IV, Eq.~(\ref{eta}). A simple
interpolation formula for the $\Phi ^{\prime }s$ would be
\begin{eqnarray}
\Phi_{Z}(x;0) & =1+c_{Z}x^{\eta},  \nonumber \\
\Phi_{\tau}(x;0) & =1+c_{\tau}x^{1-\eta}
\end{eqnarray}
Near the critical point and in the quantum disordered regime, at $T=0$ and
non-zero but small $r$, we have for $Z$
\begin{eqnarray}
Z(\omega,T & =0;r)\propto r^{(1-z/2)\nu}\Phi_{Zr}(|\omega|r^{-\nu
z};r\xi^{1/\nu}=1),  \nonumber \\
& \propto r^{(1-z/2)\nu}
\end{eqnarray}
and for the relaxation rate
\begin{eqnarray}
\tau ^{-1}(\omega ,T& =0;r)-\tau _{imp}^{-1}\propto r^{(1+z/2)\nu }\Phi
_{\tau r}(|\omega |r^{-\nu z};1)  \nonumber \\
& \propto \omega ^{2}r^{-(3z/2-1)\nu },
\end{eqnarray}%
where $\nu $ is the correlation length exponent and $z$\ is the dynamical
exponent. Here, at non-zero temperature, $|\omega |$ should be replaced by [as in Eq.~(\ref{Xscale})] $%
[\omega ^{2}+a T^{2}]^{1/2}$ , where $a$ is a constant of
order unity.

The thermopower in the limit of low temperatures, when impurity scattering
dominates, is given in terms of the conductivity $\sigma _{imp}(\mu )$ at
chemical potential $\mu $ by

\begin{equation}
S=\frac{\pi ^{2}}{3}\frac{T}{e}\frac{\partial \ln \sigma _{imp}(\mu )}{%
\partial \mu }\propto T\frac{m^{\ast }}{m}
\end{equation}%
Here we have used the fact that $\sigma _{imp}(\mu )$ does not dependent on the
renormalized effective mass, but $\mu =k_{F}^{2}/2m^{\ast }$\ such that $%
\partial \ln \sigma _{imp}/\partial \mu =(\partial \ln \sigma
_{imp}/\partial k_{F})(m^{\ast }/k_{F}).$ An experimental study of the
thermopower of YRS\cite{Hartmann10} shows a power law behavior
of $S$ in the critical regime, $S\propto T^{3/4}$,  just as predicted by our theory 
\cite{pnas}. In the Fermi liquid regime, where our theory
predicts $S/T\propto |H-H_{c}|^{-1/3}$ , experiment does show a somewhat
weaker increase of $S/T$ as $H\rightarrow H_{c}\,$.\cite%
{Hartmann10}

The nuclear spin relaxation rate is governed by the local electronic spin fluctuation spectrum:
\begin{eqnarray}
\frac{1}{T_{1}T} &\propto &\left[ \frac{1}{\omega }\int (dq){\rm Im}\chi
(q,\omega )\right] _{\omega \rightarrow 0} \\
&\propto &\frac{1}{Z^{2}(\omega =0)}\xi ^{4-d}  \nonumber
\end{eqnarray}
where  $\xi ^{-1}\propto T^{1/z}+c_{\xi }|H-H_{c}|^{\nu \eta }$ . In the
critical regime one finds $1/T_{1}T\propto T^{-(d+6)/4d}$ , and in the Fermi
liquid regime we get $1/T_{1}T\propto |H-H_{c}|^{-\nu \eta (5d+12)/3}.$
An
experimental study of the nuclear spin relaxation rate\cite%
{Ishida02} in YRS shows $1/T_{1}T\propto T^{-\alpha _{N}}$ , with $\alpha
_{N} > 1/2$. The lowest field in this experiment is about twice the critical field $H_c \sim 0.06\,{\rm T}$ so outside the critical region. Our theory prediction is $1/T_{1}T\propto T^{-3/4}$.\cite{nmr}  A more recent NMR study of YRS using magnetic field oriented parallel to the c-axis revealed a puzzling two-component signal \cite{npnmr}, one  component showing critical behavior, while the other  has Fermi liquid characteristics.
\section{Summary and Conclusion}

We have reviewed the theory of critical quasiparticles in the context of the properties of  a metallic system in the neighborhood of an antiferromagnetic quantum critical point. We discussed the conditions under which a coherent quasiparticle description may be used to analyze the behavior of metallic electrons in the critical region where their effective mass diverges as a consequence of the interaction of the fermions with the bosonic critical quantum fluctuations. The essential ingredient of the theory is accounting for the fact  that due to this interaction, both the fermions and the bosons become critical; this  leads to self-consistent conditions for the electron self energy and various vertex amplitudes that renormalize the interaction between the fermions and the critical fluctuations.

The self-consistent equations have a ``weak-coupling" solution that reproduces the conventional picture established decades ago in works by Hertz, Millis, Moriya and reviewed in various places. However there are also strong-coupling solutions in the low temperature (or excitation frequency) regime that lead to new power laws for thermodynamic and transport properties in the critical region.

When the ordered phase has a spatial variation characterized by an ordering wave vector ${\bf Q}$, as is the case for charge or spin density wave order, then there may be special regions of the Fermi surface that are connected by order parameter fluctuations of wave vector  ${\bf Q}$. These are the so-called ``hot spots". In  the general case of incommensurate antiferromagnetic quantum criticality,  the interaction of  the fermions with a single  antiferromagnetic spin fluctuation (``one-loop" analysis) leads to a highly anisotropic electron self energy that has singular behavior only at the hot regions of the Fermi surface. However, the coupling to two spin fluctuations (``two-loop" analysis), which amounts to a critical  energy fluctuation, may involve a total wave vector transfer ${\bf q}\approx 0$; this gives singular behavior over the entire Fermi surface.  This is the situation we have considered in the present work. Nevertheless, as a consequence of the internal structure of the interaction vertex of fermions with the energy fluctuation bosons, the hot and cold regions of the Fermi surface behave differently. We have analyzed the extent of the hot regions and we have shown that in the strong-coupling regime, there are no coherent critical quasiparticles in that part of the Fermi surface.  The contributions of the incoherent parts of the electron spectrum to thermodynamics and transport are expected to be small and non-singular, so we have concentrated on the effects of the singular critical quasiparticles that live at the cold regions.  

From an examination of the self-consistent expressions for the cold quasiparticles, it is evident that there is different scaling behavior in the strong  and weak coupling limits. We made this explicit by a  renormalization group analysis for the flow of an appropriately chosen ``coupling  constant" that is determined by the self energy of the cold quasiparticles.

The derived properties of the critical quasiparticles in the strong-coupling regime determine the critical behavior of various transport and thermodynamic quantities. In preparation for the discussion of the experimental consequences of our theory of critical quasiparticles, we extracted, from the self-consistent analysis, the dynamical critical exponent $z$ and the correlation length exponent $\nu$ of the critical bosonic (spin) fluctuations. These enter the expressions for the spin fluctuation spectrum and the free energy in the quantum critical region; we reviewed the  corresponding dynamical scaling form for both. From the free energy, we derived the critical behavior of thermodynamic quantities: the specific heat, the magnetization, the magnetic susceptibility, and the magnetic  Gr{\"u}neisen ratio. We reviewed the comparison of  our theoretical results for these  and also the spin fluctuation spectrum (as measured by inelastic neutron scattering) with the available data on two heavy-fermion compounds, YbRh$_{2}$Si$_{2}$ (YRS) and CeCu$_{6-x}$Au$_{x}$ at $x=0.1$ (CCA), both of which exhibit quantum critical behavior.

From the scaling behavior of the electron self energy, we gave the critical behavior of transport quantities: resistivity, magnetoresistivity, dynamical conductivity,  thermopower, and NMR relaxation rate. Where data for YRS and/or CCA 
are available, the theory gives an excellent description. All of the comparisons of theory and experiment for YRS and CCA are described and illustrated with figures in Refs.~\cite{qp1,pnas} (YRS) and Ref.~\cite{asw} (CCA).

\ack
We acknowledge useful discussions with H. v. L{\"o}hneysen, P. Gegenwart, J.D. Thompson, F. Ronning, Q. Si, C. M. Varma, and especially A.V. Chubukov and R. Fernandes. Part
of this work was performed during the summers of 2014-16
at the Aspen Center for Physics, which is supported by NSF
Grant No. PHY-1066293. J.S. and P.W. acknowledge financial
support by the Deutsche Forschungsgemeinschaft through
Grant No. SCHM 1031/4-1.
\%end{acknowledgments}
\appendix

\section{Higher order contributions to $%
\Sigma $}
In order to probe the validity of the  two-loop result for the self energy of Sec.~ IIA2, we
now estimate two higher order diagrams, containing two energy fluctuation
propagators, in parallel and crossed:

\begin{eqnarray}
\mathrm{Im}\Sigma ^{2E}(k) &=&\lambda _{2E}^{2}\int (dq_{1})\int
(dq_{2})G^{t}(k+q_{1})G^{\overline{t}}(k-q_{2}) \\
&&\times \mathrm{Im}G(k+q_{1}-q_{2})\mathrm{Im}\chi _{E}(q_{1})\mathrm{Im}%
\chi _{E}(q_{2})  \nonumber \\
&&\times \{b(\nu _{1})(1+b(\nu _{2}))(1-f(\omega +\nu _{1}-\nu _{2}))+ 
\nonumber \\
&&+(1+b(\nu _{1}))b(\nu _{2})f(\omega +\nu _{1}-\nu _{2})\}  \nonumber
\end{eqnarray}%
Here $G^{t},G^{\overline{t}}$ are the time-ordered and anti-time-ordered
Green's functions and $\lambda _{2E}=\Lambda _{v}^{3}\Lambda _{Q}^{4}$,
where an additional factor of $\Lambda _{v}\propto Z^{-1}$ arises from a
vertex correction straddling the two endpoints of $\chi _{E}$ on each side
of the diagram. The angular integrations of $G^{t},G^{\overline{t}}$ produce
factors of $1/q_{1,2}$. Using $\Lambda _{Q}\propto Z^{-1},$ we find
\begin{eqnarray}
\mathrm{Im}\Sigma ^{2E}(\mathbf{k},\omega ) &=&Z^{-14}\int_{0}^{\omega }d\nu
_{1}\int_{0}^{\nu _{1}}d\nu _{2}\int dq_{2}q_{2}^{d-2} \\
&&\times \mathrm{Im}\chi _{E}(\mathbf{q}_{2},\nu _{2})\int dq_{1}q_{1}^{d-2}%
\mathrm{Im}\chi _{E}(\mathbf{q}_{1},\nu _{1})
\end{eqnarray}%
The $q$-integrals may be done to give
\begin{equation}
\int dq_{1}q_{1}^{d-2}Z^{-6}\mathrm{Im}\chi _{E}(\mathbf{q}_{1},\nu
_{1})\propto \frac{\nu _{1}^{d-3/2}}{Z^{2d+1}}
\end{equation}%
Finally the frequency integrals may be performed with the result
\begin{equation}
\mathrm{Im}\Sigma ^{2E}(\mathbf{k},\omega )\propto \frac{\omega ^{2d-1}}{%
Z^{4(d+1)}}\propto \omega ^{2d-1-4(d+1)\eta}
\end{equation}%
which, compared with the one-loop result $\omega ^{1-\eta }$ is seen to be
smaller by a factor $\omega ^{1-3\eta }$ . We recall that $1-3\eta >0$ for
dimensions $3/2<d<9/2$\ , which means that the higher terms are irrelevant
in dimensions $d=2,3$.

\section{Vertex corrections from single
spin/energy fluctuation exchange}

\subsection{Single spin fluctuation diagram}

The relevant diagrams involve spin fluctuation propagators
connected by single-particle Green's functions in such a way that the total
momentum is\textbf{\ }$\mathbf{Q}$. The simplest diagram is the one with a
single $\chi (\mathbf{q},\nu )$ crossing as in Fig.~\ref{V} with $\Gamma$ replaced by $\chi$.
\begin{eqnarray}
\Lambda _{Q,h}^{(a)}(\mathbf{k}_{h},\omega ;\mathbf{Q,}\Omega &\rightarrow
&0)=u^{2}\Lambda _{Q}^{2}\int (dq)G(k+q-Q)  \nonumber \\
&&\times G(k+q)\chi (q+Q)  \nonumber \\
&\propto &\Sigma ^{(a)}(\mathbf{k}_{h},\omega )/\epsilon _{F},
\end{eqnarray}%
where we used $G(\mathbf{k}_{h}\mathbf{+q-Q},\omega +\nu )\approx 1/\epsilon
_{F}$ since $\mathbf{k}_{h}\mathbf{+q-Q}$ is necessarily off-shell if $%
\mathbf{k}_{h}\mathbf{,k}_{h}\mathbf{+Q}$ are onshell (remember we consider
incommensurate order). the vertex function $\Lambda _{Q,h}^{(1a)}$ is not
singular and may be discarded.

\subsection{Single energy fluctuation diagram}
The vertex function involving a single energy fluctuation
(as in Fig.~\ref{V} with $\Gamma$ replaced by $\chi_E$) is given by
\begin{equation}
\Lambda _{Q}^{(b)}(k;Q)=\Lambda _{Q}^{4}\Lambda _{v}^{2}\int
(dq)G(k+Q+q)G(k+q)\chi _{E}(q)  \nonumber
\end{equation}%
where $k=(\mathbf{k},\omega ),$ $Q=(\mathbf{Q,}\Omega )$\ and $q=(\mathbf{q}%
,\nu )$\ . We are interested in the limit $\Omega =0$ , and in momenta $%
\mathbf{k}$ on the cold part of the Fermi surface. The momentum $\mathbf{k+Q}
$ is then far from the Fermi surface and we may approximate $G(k+Q+q)\approx
1/\epsilon _{F}$ .\ The remaining expression is identical to the one for the
self-energy. 
\begin{equation}
\Lambda _{Q}^{(b)}(k;Q) \approx \frac{1}{\epsilon _{F}}\Sigma (\omega ) \propto \Lambda _{v}^{2}\Lambda _{Q}^{2d-1}(\frac{|\omega |}{\epsilon _{F}}%
)^{d-1/2} \nonumber
\end{equation}%
Anticipating critical behavior of the self energy, $\Sigma (\omega )\propto
\omega |\omega /\epsilon _{F}|^{-\eta }$ , $\eta >0$,\ we see that $\Lambda
_{Q}^{(b)}\rightarrow 0$ in the limit $\omega \rightarrow 0$. So, again,
the vertex function $\Lambda _{Q}^{(b)}$ is not singular and we will drop it.


\section*{References}

\begin{thebibliography}{99}

\bibitem{JH} J. A. Hertz, Phys. Rev. B \textbf{14}, 1165(1976).

\bibitem{millis} A. J. Millis, Phys. Rev. B \textbf{48}, 7183 (1993).

\bibitem{hvl} H. v. L{\" o}hneysen, A. Rosch, M. Vojta, and P. W{\" o}lfle, Rev. Mod. Phys. \textbf{79}, 1015 (2007). 

\bibitem{CHN} S. Chakravarty, B.I. Halperin, and D.R. Nelson,
Phys. Rev. Lett. \textbf{60}, 1057 (1988); Phys. Rev. B \textbf{39}, 2344
(1989).

\bibitem{qp1}  P. W{\" o}lfle and E. Abrahams, Phys. Rev. B 
\textbf{84}, 041101 (2011). 

\bibitem{pnas} E. Abrahams and P. W{\" o}lfle, Proc. Natl. Aca.
Sci. \textbf{109}, 3238 (2012).

\bibitem{asw} E. Abrahams, J. Schmalian, and P. W{\" o}lfle,
Phys.\ Rev.\ B \textbf{90}, 045105 (2014). 

\bibitem{Doniach77} S. Doniach, Physica B Amsterdam \textbf{91}, 231 (1977).


London \textbf{413}, 804 (2001) ; Phys. Rev. B \textbf{68}, 115103 (2003).

\bibitem{Si10} Q. Si, Phys. Status Solidi B \textbf{247}, 476--484 (2010) and references therein.

\bibitem{Pepin07} C. P\'{e}pin, Phys. Rev. Lett. \textbf{98}, 206401 (2007);
Phys. Rev. B \textbf{77}, 245129 (2008).

\bibitem{Kotliar08} L. De Leo, M. Civelli, and G. Kotliar, Phys. Rev. B 
\textbf{77}, 075107 (2008).




\bibitem{Senthil06} T. Senthil, Ann. Phys. \textbf{321}, 1669 (2006) and references therein.

\bibitem{Sachdev08} S. Sachdev, Nat. Phys. \textbf{4}, 173 (2008). 

\bibitem{Lee08} P. A. Lee, Rep.Prog. Phys. \textbf{71}, 012501 (2008).

\bibitem{Park08} T. Park, M. J. Graf, L. Boulaevskii, J. L. Sarrao, and J.
D. Thompson, Proc. Natl. Acad. Sci. U.S.A. \textbf{105}, 6825 (2008) .

\bibitem{Shishido05} H. Shishido, R. Settai, H. Harima, and Y. Onuki, J.
Phys. Soc.

Jpn. \textbf{74}, 1103 (2005) .

\bibitem{Goh08} S. K. Goh, J. Paglione, M. Sutherland, E. C. T. O'Farrell,
C. Bergemann, T. A. Sayles, and M. B. Maple, Phys. Rev. Lett. \textbf{101}, 056402 (2008).

\bibitem{Vojta08} M. Vojta, Phys. Rev. B{\bf 78},125109 (2008).

\bibitem{gegen07}P. Gegenwart, T. Westerkamp, C. Krellner, Y. Tokiwa, S. Paschen, C. Geibel, F. Steglich, E. Abrahams, Q. Si,  Science {\bf 315}, 969 (2007).

\bibitem{Kummer15} K. Kummer, S. Patil, A. Chikina, M. G\"{u}ttler, M. H\"{o}%
ppner, A. Generalov, S. Danzenb\"{a}cher, S. Seiro, A. Hannaske, C.
Krellner, Yu. Kucherenko, M. Shi, M. Radovic, E. Rienks, G. Zwicknagl, K.
Matho, J.W. Allen, C. Laubschat, C. Geibel, and D. V. Vyalikh, Phys. Rev. X 
\textbf{5}, 011028 (2015).


\bibitem{WA15}P. W{\" o}lfle and E. Abrahams, Phys. Rev. B \textbf{92}, 155111(2015). 

\bibitem{WA16} P. W{\" o}lfle and E. Abrahams, Phys. Rev. B \textbf{93}, 075128 (2016).

\bibitem{Weiss16}  P. S. Weiss, B. N. Narozhny, J. Schmalian,  and Peter W{\"o}lfle, Phys. Rev. B \textbf{93}, 045128 (2016).

\bibitem{Abanov00} Ar. Abanov and Andrey V. Chubukov Phys. Rev. Lett. {\bf 84}, 5608  (2000).

\bibitem{Abanov03} A. Abanov, A. V. Chubukov and J. Schmalian, Adv. Phys. \textbf{52}, 119 (2003).


\bibitem{Hartnoll11} S. A. Hartnoll, D. M. Hofman, M. A. Metlitski, and S. Sachdev, Phys. Rev. B \textbf{84}, 125115 (2011).


\bibitem{Novikov83} V.A. Novikov, M.A. Shifman, A.I. Vainshtein, and V.I. Zakharov Nucl. Phys. \textbf{B229}, 381 (1983).

\bibitem{Stock12}C. Stock, C. Broholm, Y. Zhao, F. Demmel, H. J. Kang, K. C. Rule, and C. Petrovic, Phys. Rev. Lett. {\bf 109}, 167207 (2012)

\bibitem{Schroeder00} A. Schr\"{o}der, G. Aeppli, R. Coldea, M. Adams, O.
Stockert,

H.v. L\"{o}hneysen, E. Bucher, R. Ramazashvili, and P. Coleman, Nature 
\textbf{407}, 351 (2000).

\bibitem{Stockert10} O. Stockert, H. v. L\"{o}hneysen, W. Schmidt, M.
Enderle, and M. Loewenhaupt, J. Low Temp. Phys. \textbf{161}, 55 (2010).

\bibitem{Rosch97} A. Rosch, A. Schr\"{o}der, O. Stockert, and H. v. L\"{o}hneysen
Phys. Rev. Lett. {\bf 79}, 159 (1997).

\bibitem{Oeschler08} N. Oeschler, S. Hartmann, A.P. Pikul, C. Krellner, C.
Geibel, and F. Steglich, Physica B \textbf{403}, 1254 (2008).

\bibitem{hvl96} H. v. L\"{o}hneysen, M. Sieck, O. Stockert, and
M.Waffenschmidt, Physica B \textbf{223-224}, 471 (1996).

\bibitem{Custers03} J. Custers, P. Gegenwart, H. Wilhelm, K. Neumaier, Y.
Tokiwa, O. Trovarelli, C. Geibel, F. Steglich, C. P{\' e}pin, and P. Coleman, Nature \textbf{424}, 524 (2003).

\bibitem{Tokiwa09} Y. Tokiwa, T. Radu, C. Geibel, F. Steglich, P. Gegenwart,
Phys. Rev. Lett. \textbf{104,} 096401 (2009).

\bibitem{Gegenwart02} P. Gegenwart, J. Custers, C. Geibel, K. Neumaier, T.
Tayama, K. Tenya, O. Trovarelli, and F. Steglich, Phys. Rev. Lett. \textbf{89%
}, 056402 (2002).

\bibitem{Zhu03} L. Zhu, M. Garst, A. Rosch, and Q. Si, Phys. Rev. B \textbf{%
91}, 066404 (2003).

\bibitem{hvl14} H. v. L\"{o}hneysen, private communication.

\bibitem{Gegenwart08} P. Gegenwart, Q. Si, and F. Steglich, Nature Phys. 
\textbf{4}, 186 (2008).

\bibitem{Taupin15} M. Taupin, G. Knebel, T. D. Matsuda, G. Lapertot, Y.
Machida, K. Izawa, J.-P. Brison, and J. Flouquet, Phys. Rev. Lett. \textbf{%
115}, 046402 (2015).

\bibitem{hvl01} H. v. L\"{o}hneysen, C. Pfleiderer, T. Pietrus, O. Stockert,
and B. Will, Phys. Rev. B, \textbf{63}, 134411 (2001).

\bibitem{Hartmann10} S. Hartmann, N. Oeschler, C. Krellner, C. Geibel, S.
Paschen, and Frank Steglich, Phys. Rev. Lett. \textbf{104,} 096401 (2010).

\bibitem{Ishida02} K. Ishida, K. Okamoto, Y. Kawasaki, Y. Kitaoka, O.
Trovarelli, C. Geibel, and F. Steglich, Phys. Rev. Lett. \textbf{89}, 107202
(2002).
\bibitem{nmr} This was already disciussed in Ref.~\cite{qp1}, where the result was given in terms of an unknown exponent $\beta$, which our scaling analysis now determines as $\beta=-1/4$. Hence the result reported here. 

\bibitem{npnmr}  S. Kambe, H. Saka,  Y. Tokunaga, G. Lapertot, T. D. Matsuda, G. Knebel, J. Flouquet
and R. E. Walstedt, Nat. Phys. {\bf 10}, 840 (2014).

\end{thebibliography}
\end{document}